\documentclass[preprint,12pt,sort&compress,times]{elsarticle}

\usepackage{graphicx}
\usepackage{verbatim} 
\usepackage{amsmath}
\usepackage{amssymb}
\usepackage[usenames]{color}
\usepackage{xspace}

\definecolor{DarkRed}{rgb}{0.6,0,0}
\definecolor{DarkBlue}{rgb}{0,0,0.6} 

\newcommand{\vect}[1]{\mathbf{#1}}
\newcommand{\units}[1]{\,\mathrm{#1}}
\newcommand{\dexp}[1]{\cdot 10^{#1}}
\newcommand{\dpartial}[2]{\frac{\partial #1}{\partial #2}}
\newcommand{\parref}[1]{(\ref{#1})}

\newcommand{\net}{\tilde{n_e}}

\newcommand{\iaa}{
IAA-CSIC, P.O. Box 3004, 18080 Granada, Spain}

\journal{Journal of Computational Physics}

\begin{document}
\begin{frontmatter}

\title{Density models for streamer discharges:\\ beyond cylindrical symmetry and homogeneous media}

\author{A Luque}
\address{\iaa}
\ead{luque@iaa.es}
\author{U Ebert}
\address{Centrum Wiskunde \& Informatica (CWI), Amsterdam, The Netherlands,\\
and Department of Applied Physics, Technische Universiteit Eindhoven, The Netherlands}

\begin{abstract}
Streamer electrical discharges are often investigated with computer simulations of density models (also called drift-diffusion-reaction models).
We review these models, detailing their physical foundations, their range of validity and the most relevant numerical algorithms employed in solving them.  We focus particularly on schemes of adaptative refinement, used to resolve the multiple length scales in a streamer discharge without a high computational cost.
We then report recent results from these models, emphasizing developments that go beyond cylindrically symmetrical streamers propagating in homogeneous media.
\end{abstract}

\end{frontmatter}

\section{Introduction}
Streamers \cite{Raether1939/ZPhy,Raizer1991/book,Ebert2006/PSST,Ebert2008/JPhD} are transient electrical discharges that appear when a non-conducting medium is suddenly exposed to a high electric potential. While the average background field might be too low for plasma generation by electron impact on neutral molecules, the streamer discharge channel can enhance the electric field at its growing tip so strongly that it can create additional plasma and propagate nevertheless. The streamer achieves this through a very nonlinear dynamics with an intricate inner structure and locally very steep density gradients. This structure generates the local field enhancement and maintains propagation. Therefore, the accurate simulation of single, cylindrically symmetric streamer channels is far from trivial, even if the electrons are approximated by their densities as is conventionally done.

It should be noted that the full process evolves on even more scales. On the one hand, when the electric field ahead of the streamer head is very high, single electrons can run away and generate hard radiation through Bremsstrahlung. To describe these effects, single electrons have to be followed in a particle model; numerical methods to track these particles in an efficient manner and the derivation of density models are discussed in this special issue by Li {\it et al.}~\cite{Li2011/JCP}. On the other hand, in the laboratory, in technological applications and in lightning related processes, hundreds to ten-thousands of streamer channels can propagate next to each other. In this case it is vital to develop models on a more macroscopic scale than the density approximation. Through model reduction, one can derive moving boundary models \cite{Lozansky1973/JPhD,Meulenbroek2004/PhRvE, Brau2008/PhRvE/1,Brau2008/PhRvE,Brau2009/PhRvE} for the underlying ionization fronts, or one even can try to develop models for the streamer channel as a whole.

Within the current short review, we discuss the justification for density models for streamers, the numerical solution strategy, and then some results that go beyond single cylindrically symmetric streamers in homogeneous media. We treat interacting streamers, first through the trick of considering an infinite array of identical streamers next to each other, then through full 3D simulations. Next, we discuss streamer branching in full 3D.  Finally we discuss how to simulate the emergence of sprite streamers from earth's ionosphere; a peculiar issue is here how to combine the very different scales of the electric fields generated by the thundercloud-ground-ionosphere system with the fine inner structure of the discharge and the varying density of the atmosphere. 

With the density models, we here focus on the oldest and most extensively investigated family of numerical streamer models; they are also known as fluid models, continuous models or reaction-drift-diffusion models.  These models are now mature enough that their predictions often deviate from experimental observation by a margin comparable to the random variations and uncertainties of the experiments. 

\section{Model formulation}
\label{formulation}

\subsection{The density model}

A classical density model for a streamer discharge has the structure of a reaction-drift-diffusion equation for the electrons and reaction equations for various ions and excited species coupled to the electric field
\begin{subequations}
\begin{eqnarray}
 \label{fluid}
  \dpartial{n_e}{t} & = &
   \nabla \cdot (n_e \mu_e \vect{E}) + \nabla \cdot D_e \nabla n_e + S^{im}_e + S^{ph}_e,  \label{sigma}\\
  \dpartial{[Z_i]}{t} & = & S^{im}_i + S^{ph}_i, ~~~i=1,\ldots,N\label{endfluid},\\
\epsilon_0 \nabla\cdot\vect{E}& =& q,~~~\vect{E}=-\nabla\phi.  \label{field}
\end{eqnarray}
\end{subequations}
Here $n_e$ is the electron number density and $[Z_i]$ is the density of the heavy species $Z_i$ denoting an ion or an excited state, and $\mu_e$ and $D_e$ are mobility and diffusion coefficient of the electrons. $S^{im}_{e,i}$ denotes mostly local generation or loss of species due to reaction at direct encounter of particles; the most prominent example is electron impact ionization, i.e., the generation of electron-ion pairs through impact of electrons on neutrals; the efficiency of this process strongly depends on the electron energy that in turn is determined by gas density and electric field. $S^{ph}_{e,i}$ denotes the generation of species through radiative transport in a generically nonlocal process; the most prominent example in streamer physics is the generation of electron-O$_2^-$ pairs through photo-ionization in air. Finally, $\vect{E}$ is the electric field, $\phi$ the electric potential, and $q$ is the local space charge, determined as $q=\sum_i q_i\;[Z_i]-{\rm e}\;n_e$; here e is the elementary charge and $q_i=-{\rm e},0,+{\rm e}$ is the charge of species $Z_i$.

\subsection{The underlying approximations}

We list here the main
approximations underlying this model. For a more through derivation of the model, we refer to \cite{Li2011/JCP,Li2010/JCoPh,Li2007/JAP}.  

1. The electrons are much more mobile than the ions, and the dynamics takes place on their time scale, therefore ion motion is neglected --- whether the evaluation time scales exceed this limit, needs to be checked after evaluations. 

2. The electron motion is approximated by drift and diffusion within the local field. This entails that the electrons rapidly relax to a velocity where the acceleration by the electric field exactly cancels the momentum losses due to collisions with other particles, and that spatial or temporal variations of the electric field are not important. The accuracy of the drift-diffusion approximation and the local field approximation for electron currents in streamer ionization fronts was verified in~\cite{Li2007/JAP} for electric fields up to some threshold; this statement holds for arbitrary gas density. However, when the field at the streamer head exceeds a limit (180 kV/cm in air at standard temperature and pressure, STP~\cite{Li2011/JCP,Li2009/JPhD,Chanrion2010/JGR,Li2010/JCoPh}), some electrons in
the tail of the electron energy distribution function might not relax to a steady velocity anymore, but run away with a field dependent probability. Above the thermal runaway electric field (260 kV/cm in STP air \cite{Phelps1987/PhRvA,Moss2006/JGRA,Vrhovac2001/JAP}) the bulk of the electrons will keep accelerating up to relativistic energies and the drift approximation breaks down.

3. The field is calculated in electrostatic approximation. For typical current densities and diameters of streamers~\cite{Ebert2010/JGRA}, this approximation is justified.

4. The ensembles of electrons and ions can be treated in density approximation. The validity of this approximation depends first on the gas density, and second on the specific region within the streamer.  First, streamers in different gas densities $n$ are related through similarity relations or Townsend scaling in very good approximation~\cite{Pasko1998b/GeoRL,Ebert2010/JGRA}; these relations imply that the electron density $n_e$ in a streamer scales as $n^2$, and that the intrinsic length scales scale as $1/n$. Consequently, the total number of electrons in similar parts of streamers scales as $1/n$. This implies that the density approximation in all regions of the streamer becomes better when the gas density decreases. On the contrary, the density approximation for electrons and ions becomes worse when the density of the neutral medium increases, and it has to be reconsidered at the high densities of liquids. (Other approximations to be reconsidered in streamers at liquid densities are the absence of heating and the neglect of electron-electron and electron-ion collisions in the source terms $S_{im}$; in gases only electron-neutral collisions are included in typical streamer models, but the degree of ionization and therefore the probability of electron-electron or electron-ion collisions increases with increasing density because $n_e/n \propto n$.) A review of validity and corrections to these similarity solutions is given in~\cite{Ebert2010/JGRA}. 

Second, different spatial regions of the streamer have to be distinguished: interior, front and outer region. The streamer interior has large densities of electrons and ions and little inner structure; the density approximation is certainly justified there. In the outer region, electron densities are typically very low (except for high discharge repetition frequencies or with other strong ionization sources) and densities should rather be interpreted as probabilities; however, a change of this low background ionization has essentially no influence on streamer propagation in typical situations~\cite{Wormeester2010/arxiv}, but it can have an influence on branching and "feather" formation in very pure gases with little nonlocal photo-ionization~\cite{Nijdam2010/JPD,Wormeester2010/arxiv}. Finally, the streamer ionization front connects interior and exterior and the electron density can develop steep gradients, in particular, in high electric fields and with little or no nonlocal photo-ionization~\cite{Montijn2006/JCoPh, Li2007/JAP,Nijdam2010/JPD,Wormeester2010/arxiv}. Within this region, the local density approximation can break down as electrons with a higher energy are ahead of others. The electron energies are then determined not only by the local electric field, but also by the gradient of the electron density; this effect can be incorporated in a modification of the impact ionization term; or alternatively, the model can be extended by an electron energy equation~\cite{Li2010/JCoPh}. These terms can be included in the general structure of the reaction-drift-diffusion model above and do not require any separate numerical strategy. According to~\cite{Li2010/JCoPh}, the density approximation describes streamer propagation well, as long as no massive electron runaway sets in. However, to properly describe the effect of density fluctuations and electron run-away on streamer stability, a particle approach should be chosen. Presently available models are Monte Carlo particle models~\cite{Moss2006/JGRA, Chanrion2008/JCoPh, Soria-Hoyo2009/JCoPh} and spatially hybrid models \cite{Li2010/JCoPh,Li2011/JCP} that combine particle and density models in different volumes of the simulation.

\subsection{Typical models in air}

The general structure of the density model defined above has been recognized decades ago but the search for better reaction and transport parameters in different gases goes on. Most modeling work is devoted to air modeled as a mixture of nitrogen and oxygen.

The impact ionization term $S^{im}$ in air includes the generation of free electrons and positive ions through electron impact on neutral molecules as well as electron losses due to attachment to oxygen. In particular the first reaction strongly depends on the electric field. There is also field dependent detachment of electrons from O$_2^-$~\cite{Pancheshnyi2005/PSST}, however, it is typically neglected in air, and it indeed only contributes in a relevant manner in nitrogen with admixtures of oxygen as low as 1 p.p.m., but not at the 20\% oxygen content in air~\cite{Wormeester2010/arxiv}.

This is because air has very strong nonlocal photo-ionization --- for a historical review of this concept, we refer to the introduction of~\cite{Nijdam2010/JPD}. In oxygen-nitrogen mixtures such as air, photo-ionization is in most cases included through the Zhelezniak model~\cite{Zhelezniak1982/TepVT}.  In this model, the states $\mathrm{b^1\Pi_u}$, $\mathrm{b'^1\Sigma_u^+}$ and $\mathrm{c'^1_4\Sigma_u^+}$ of molecular nitrogen, populated by impact excitation in the high-field regions of the streamer, decay by emitting photons, some with energies above the ionization threshold of molecular oxygen, 12.1 eV ($1025\,\mathrm{\AA}$). 
The Zhelezniak model assumes that the excitation of the emitting states is roughly proportional to the impact ionization.  Then
the production of electron-ion pairs per unit of volume and time is written as
\begin{equation}
  \newcommand{\R}{\vect{r}}
  \label{Sph_int_dim}
  S_{ph}(\vect{\R}) = \frac{\xi}{4\pi} \frac{p_q}{p+p_q}. \int
  \frac{h(p|\vect{\R} - \vect{\R}'|)
          S^{im}_e(\vect{\R}')d^3(p\vect{\R}')}
       {|p\vect{\R} - p\vect{\R}'|^2},
\end{equation}
where $\xi$ is a proportionality constant, $p_q=60 \units{Torr}=80 \units{mbar}$ is the quenching pressure of the emitting states and $h$ is the absorption function of photo-ionizing radiation. Photo-ionization might also be very strong in hydrogen-helium mixtures that dominate the atmospheres of gas giants like Jupiter and Saturn~\cite{Dubrovin2010/JGR}; however, quantitative models are not available for this mixture. 

The reaction rates strongly depend on electron energies (that are typically parameterized by the electric field that accelerates the electrons), but also the transport coefficients $\mu_e$ and $D_e$ depend on it, though weakly. In most literature, they are treated as constant. Furthermore, in general the diffusion coefficient $D_e$ is a tensor with different components parallel and perpendicular to the electric field~\cite{Li2011/JCP}, but most investigators approximate it as a scalar.
Reaction and transport coefficients are mostly calculated in advance from cross-section data such as~\cite{Phelps1985/PhRvA} by solving the Boltzmann equation for electrons with numerical codes such as BOLSIG+ \cite{Hagelaar2005/PSST}. Alternatively, they also can be calculated by performing Monte Carlo simulations with good statistics~\cite{Li2010/JCoPh,Li2011/JCP}. Much work is currently devoted to improvements of the cross-section data as well as to the evaluation of the Boltzmann equation \cite{Dujko2010/POAB}.

On the other hand, it can be stated that streamer discharges seem to be amazingly robust both against model changes in simulations~\cite{Wormeester2010/arxiv} as well as against changes of gas composition in experiments~\cite{Nijdam2010/JPD,Dubrovin2010/JGR}. The reason is probably the strongly nonlinear dynamics of the streamer that approaches generic attractors of the dynamics independently of microscopic details. E.g., in streamers in nitrogen with 1 p.p.m. oxygen, the effect of background ionization or photo-ionization can hardly be distinguished, and major changes of the photo-ionization model in air have only minor effects on streamer velocity~\cite{Wormeester2010/arxiv}.

\section{Numerical schemes}
\label{numerics}
Generally speaking, the physical models that underlie numerical streamer simulations are always small variations on the model discussed above.  These variations include e.g. the use of field-dependent mobility and diffusion and the inclusion of kinetic equations for additional chemical reactions. 

To reduce computational demand, almost all streamer
simulations are performed in cylindrical symmetry with
radial and longitudinal coordinates $(r,z)$ although below we will discuss some recent extensions to real 3d.

\subsection{Discretization of the transport equation}
The main challenge in the numerical solution of the densities of charged particles in streamers arises from the very steep gradients in the streamer tip.  The choice of discretizations has been largely driven by the requirement of handling these gradients efficiently. The first studies~\cite{Dhali1985/PhRvA, Dhali1987/JAP, Wu1988/PhRvA,Vitello1994/PhRvE} used variations on the Flux-Corrected Transport (FCT) algorithm developed by Zalesak~\cite{Zalesak1979/JCoPh}, combined with a modified Euler scheme for the time integration.   In 1995 Kulikovsky~\cite{Kulikovsky1995/JCoPh} introduced a scheme based on improvements of the Scharfetter-Gummel algorithm that has subsequently been used also by Liu and Pasko~\cite{Liu2004/JGRA/1} and Bourdon and coworkers~\cite{Bourdon2007/PSST}.  On the other hand, first-order upwind schemes were used in~\cite{Pancheshnyi2003/JPhD} but they seemed to be too diffusive, and they over-stabilized the streamer fronts. In~\cite{Montijn2006/JCoPh}, an upwind-biased scheme was applied that used the Koren flux limiter \cite{Koren1993/inbook} to switch between first-order upwind and third order upwind-biased.

These discretizations may be combined with various algorithms to work
with different resolutions in different volumes of the domain. The
importance of this part of the numerical scheme stems from the
multi-scale nature of the streamer process discussed
above. Kulikovsky~\cite{Kulikovsky2000/JPhD} used a ``window'' with a
fixed, higher resolution that followed the streamer head, while
Pancheshnyi and coauthors~\cite{Pancheshnyi2005/PhRvE} and Eichwald
and coauthors~\cite{Eichwald2008/JPhD} used a rectangular product grid
with smoothly varying space steps.  Min and coauthors \cite{Min2001/ITM} and
Nikandrov and coauthors \cite{Nikandrov2008/ITPS} have used dynamically adaptive meshes.

Montijn and coauthors~\cite{Montijn2006/JCoPh} used a scheme of nested
uniform grids generated adaptively with a criterium based on the
second derivatives of the species and charge densities as follows.  
The refinement builds a hierarchy of grids starting from the coarsest grid that covers the entire volume and descending to finer grids.  To decide the
areas to refine, at each grid at level $l$, monitor functions $M_u$ are calculated, where $u$ stands for the electron density or space charge density.  These functions come from the discretization of
\begin{equation}
  M_u(r, z) = (\Delta r^l)^2 \left|
    \frac{1}{r} \frac{\partial}{\partial r} 
      \left( r \frac{\partial u}{\partial r}\right)\right|
    + (\Delta z^l)^2 \left|\frac{\partial^2 u}{\partial z^2}\right|,
\end{equation}
where $\Delta r^l$ and $\Delta z^l$ are, respectively, the $l$-level grid
resolutions in $r$ and $z$.  The refinement criterium is then
specified from grid-independent refinement tolerances $\epsilon_u$ as
\begin{equation}
  \label{refinement}
  \text{refine all grid cells $i, j$ where }\;
     \frac{M_u(r_i^l, z_j^l)}{\max u_{ij}^l} \ge \epsilon_u.
\end{equation}
Streamer fronts are pulled, i.e. the dynamics is determined in the leading edge of the ionization front \cite{Ebert2000/PhyD}.  For curved fronts as occuring in streamers, the leading edge is limited by the decay of the electric field.  Therefore
the refinement criterium \parref{refinement} is often insufficient in the leading
edge of the front.  The approach used in \cite{Montijn2006/JCoPh} is
to use a region larger than given by \parref{refinement} and extend all grids in the leading edge up to the
layer where the electron density is below a small threshold.  However,
when photo-ionization is included in the simulations the electron
density ahead of the front decays much slower and this approach leads to
an overextension of the refinement grids.  Instead one can use a
criterium based on the absolute value of the electric field, adding to 
\parref{refinement} the additional criterium
\begin{equation}
  \label{refinement_eabs}
  \text{refine grid cells $i, j$ where }\;
    |E_{ij}| \ge E_{threshold},
\end{equation}
with an appropiately choosen $E_{threshold}$.

\subsection{Strategies for the Poisson-equation}
Solving the Poisson equation is usually one of the most expensive
parts of simulations of density models for streamers.  It is
challenging, because the solution is nonlocal and instantaneous, i.e.,
the boundary conditions together with the very localized charge
distribution in the simulation domain determine the electric field
instantaneously.  The Poisson equation is
most often discretized in a cartesian grid by a five-point second
order scheme. The resulting linear system is then solved using either
iterative methods, usually from the Succesive Over-Relaxation (SOR) 
family~\cite{Kulikovsky1997/JPhD/1, Pancheshnyi2001/JPhD/1,
  Liu2004/JGRA/1, Eichwald2008/JPhD} or by direct methods such as
cyclic reduction~\cite{Montijn2006/JCoPh, Wackers2005/JCoAM} and
SuperLU~\cite{Celestin2009/JPhD}. In general, direct methods can be
much faster than iterative methods for these problems, but they
perform similarly if the iterative method is provided with
a good initial guess from the solution of a previous timestep.

Even with fast algorithms, solving the Poisson equation is a major
computational bottleneck for streamer simulations in full three
dimensions.  Two earlier 3D streamer
simulations~\cite{Kulikovsky1998/PhLA,Pancheshnyi2005/PSST} constitute
a proof of principle that such computations can be performed. An
efficient and parallelizable approach for that problem is described
in~\cite{Luque2008/PhRvL}; it extends the grid refinement technique
from~\cite{Montijn2006/JCoPh} to 3D simulations.  This method is
efficient for streamer dynamics that can be represented in
cylindrical coordinates $(r, z, \theta)$ with a relatively low
resolution in the axial coordinate $\Delta\theta = 2\pi / N$.

The discrete Fourier transform (DFT) in $\theta$ of the electrostatic
potential $\phi(r, z, \theta)$ reads
\begin{equation}
  \tilde{\phi}_k(r, z) = \sum_{n=0}^{N-1} \phi(r, z, 2\pi n / N) 
  e^{-2\pi ikn/N}, \label{dft}
\end{equation}
which implies the usual properties of the DFT of a real quantity with
$N$ even,
\begin{equation}
  \tilde{\phi}_k(r, z) = \tilde{\phi}^\star_{N-k}(r, z), \;
  \tilde{\phi}_0(r, z) = \tilde{\phi}^\star_0(r, z), \;
  \tilde{\phi}_{N/2}(r, z) = \tilde{\phi}^\star_{N/2}(r, z),
  \label{dft-real}
\end{equation}
where $^\star$ indicates complex conjugation.  The benefit of working
in Fourier space is that different modes $k$ are decoupled and can be
solved in parallel.  The Laplace operator $\nabla^2$ can be decomposed
into $\nabla^2 = \nabla^2_{rz} + \frac{1}{r^2} 
\frac{\partial^2}{\partial \theta^2}$ and the midpoint discretization of
the second derivative with respect to $\theta$ is
\begin{equation}
  D^2_\theta \phi_j = \frac{\phi_{j-1} - 2 \phi_j + \phi_{j+1}}{\Delta
    \theta^2},
\end{equation}
where for clarity we have omitted the indices for $r$ and $z$.
Applying the DFT as defined in \parref{dft} we obtain
\begin{equation}
  D^2_\theta \tilde{\phi}_k = 
    \frac{\tilde{\phi}_ke^{-2\pi i k/N} -  2 \tilde{\phi}_k + \tilde{\phi}_ke^{2\pi i
      k/N}}{\Delta \theta^2} = -|w_k|^2 \tilde{\phi}_k,
\end{equation}
where 
\begin{equation}
  |w_k|^2 = \frac{2}{\Delta \theta^2}(1 - \cos k \Delta \theta).
\end{equation}
Note that as $\Delta \theta \to 0$, $|w_k|^2 \to k^2$, giving the well
known form of the continuous Fourier transform of a second derivative.

Finally, the equation for $\tilde{\phi}$ including the $r, z$ terms is
a two-dimensional Helmholtz equation
\begin{equation}
  [\nabla_{rz}^2] \tilde{\phi}_k(r, z) - \frac{|w_k|^2}{r^2}
  \tilde{\phi}_k(r, z) = -\tilde{q}_k(r, z) / \epsilon_0,
  \label{poisson-dft}
\end{equation}
where $[\nabla_{rz}^2]$ is the discrete Laplace operator in $(r, z)$
and $\tilde{q}_k(r, z)$ is the DFT of the space charge density.
The available numerical methods for this equation are basically the
same as for the Poisson equation (where the second term is missing)
but \parref{poisson-dft} makes it possible to solve the equations
independently and hence in parallel for each $k$.

Unfortunately, the reaction terms in \parref{fluid}-\parref{endfluid}
are nonlinear so the convection-diffusion-reaction side of the model
must be solved in real space, where it can also be trivially
parallelized.  Hence an efficient parallel algorithm must switch
between Fourier and real space at each timestep.  This is relatively
fast by using widely available Fast Fourier Transform (FFT) codes.

Since the Poisson equation is usually solved in a cartesian grid, a
second problem concerns the modeling of a curvilinear or point-shaped
electrode. This is an important aspect of streamer simulations since
in experiments streamers almost always emerge from a pointed
electrode. One common approach, introduced by Babaeva and
Naidis~\cite{Babaeva1996/JPhD}, is to consider a spherical needle electrode where the computational domain of the densities begins only at the tip of the electrode, and to calculate its influence by the method of image charges. Another approach was proposed in~\cite{Luque2008/JPhD}, based on a simplified version of the method of virtual charges. A similar approach that does include the needle electrode into the domain of density computations is the Charge Simulation Method \cite{Singer1974/IEEE, Malik1989/IEEE} that was implemented for streamers in section 7.2 of \cite{Li2009/PhDT}.
Recently the use of immersed boundaries was
introduced in streamer simulations by Celestin {\it et al}.~\cite{Celestin2009/JPhD}.

A non-planar electrode can be also simulated by using curvilinear coordinates, adapted to the geometrical shape of the electrodes 
\cite{Serdyuk2001/JPhD,van_Dijk2009/JPhD}; this is particularly appropriate if the pointed-electrode has a parabolic shape.

\subsection{Strategies for the nonlocal photo-ionization}

In the first simulations of streamers, photo-ionization was not included in the model but rather substituted by an unrealistically high constant pre-ionization~\cite{Dhali1985/PhRvA}.  Later, the Zhelezniak model was introduced by directly evaluating the integral (\ref{Sph_int_dim}), i.e., by summing the effect of each emitting source on each grid point in the simulation domain.  The number of computations needed in this approach scales quadratically with the total number of grid points, growing much faster than the number of computations required for the Poisson equation or the advection-diffusion time stepping. One can use a coarser grid for photo-ionization calculations~\cite{Kulikovsky2000/JPhD}, but this seems to introduce spurious effects.  Another approach, used in \cite{Steinle1999/JPhD}, is to approximate the spatial distribution of the photo-ionizing emitters by a sphere centered at the center of mass of radiation and with a radius equal to some characteristic radius of the streamer.  

A more accurate approach is to approximate the Zhelezniak integral by the solution of elliptic partial differential equations.  This method was introduced independently by Segur {\it et al.}~\cite{Segur2006/PSST} and Luque {\it et al.}~\cite{Luque2007/ApPhL}, and further refined in~\cite{Liu2007/ApPhL, Bourdon2007/PSST, Capeillere2008/JPhD}.

\section{Beyond the cylindrically symmetrical streamer in 
  a homogeneous gas}
\label{results}

The keystone of streamer simulations consists of a single
cylindrically symmetrical streamer propagating in a homogeneous
medium.  As we mentioned above, this is by itself a physically relevant
and challenging numerical problem.  Nevertheless, in recent years
several extensions to this paradigm have been introduced that investigate a wider range of problems.  Here we review three of
them: the interaction between neighboring streamers, the branching of
a streamer channel and the propagation of streamers in the upper
atmosphere, with a significant gradient of air density.

\subsection{Streamer interaction}
Streamers very rarely appear in isolation; most often they belong to
bunches of many streamers, either because they have emerged at different
points of a wire electrode or because they are contained in a highly
branched tree emerging from a point electrode.  Since each streamer
carries charge, they interact electrostatically.

\subsubsection{Bunches of streamers and Saffman-Taylor streamers}
The first work to address the interaction of neighbouring streamers was
by Naidis \cite{Naidis1996/JPhD}.  He used a quasi-2D density model to
estimate the effect of the interaction from neighbouring streamers in a
regular one-dimensional lattice of well-separated streamers.  The
underlying assumption was that the radius of each streamer is
much smaller than the distance to the closest neighbour.

\begin{figure}
\includegraphics[width=.45\textwidth]{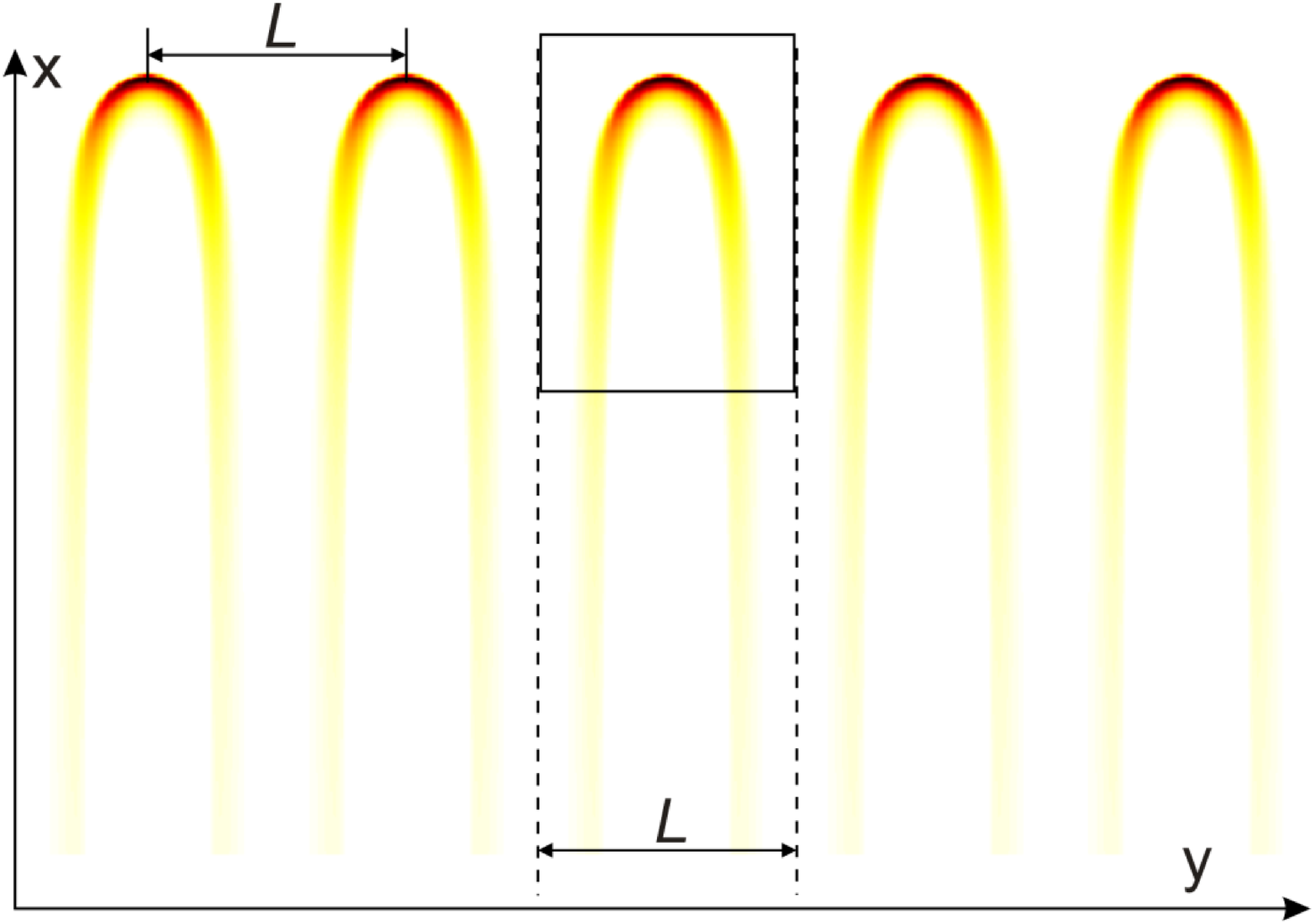} 
\includegraphics[width=.45\textwidth]{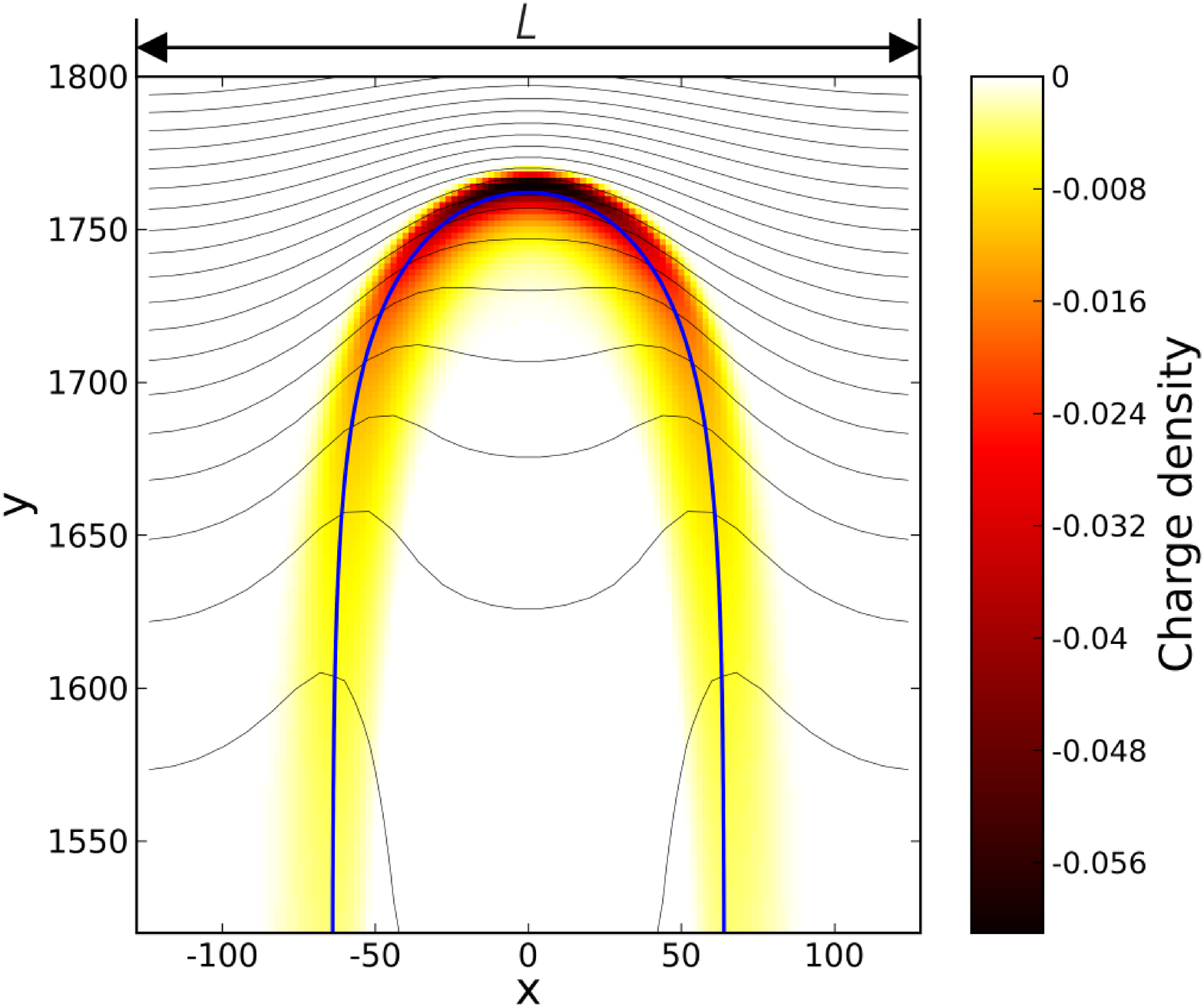} 
\caption{\label{saffman} \textbf{Left:} periodic array of negative 
streamers.  The interaction between neighboring streamers can be
incorporated into a density model by imposing homogeneous Neumann boundary
conditions on the electrostatic potential on the dashed lines. 
\textbf{Right:}  The resulting streamer profile fits the analytical
solution of a selected Saffman-Taylor finger, with width $L/2$.
(reproduced from~\cite{Luque2008/PhRvE})}
\end{figure}

A further step was demonstrated in \cite{Luque2008/PhRvE}, 
where the transversal dynamics of 
interacting streamers is taken into
account by a modification of a single-streamer model.
As shown in Fig.~\ref{saffman} (left) the interaction between 
neighbouring streamers is incorporated as a lateral homogeneous 
Neumann boundary condition of the electrostatic potential.  As a first
step, Ref.~\cite{Luque2008/PhRvE} was limited to a 2D (planar) model of negative streamers without photoionization.  However the technique can also be applied to more realistic situations \cite{Ratushna/prep}.

The outcome, after a transient regime where each
streamer expands, is a steady state where each channel fills
half of the available width $L$ (the distance between one streamer and
its neighbour).  In this steady state, the electric field far behind
the array of streamer heads is completely screened.  The numerical
solution shows a remarkable agreement with the
Saffman-Taylor analytical solution of the well-known problem of
viscous fingering in Hele-Shaw cells (Fig.~\ref{saffman}, right)
\cite{Saffman1958/RSPSA, Bensimon1986/RvMP}.

The analogy between streamers and viscous fingering is explicit in moving
boundary models \cite{Lozansky1973/JPhD, Meulenbroek2005/PhRvL, 
Brau2008/PhRvE,Brau2008/PhRvE/1,Brau2009/PhRvE}.

\subsubsection{Interaction between two streamer heads}
\begin{figure}
\includegraphics[width=.45\textwidth]{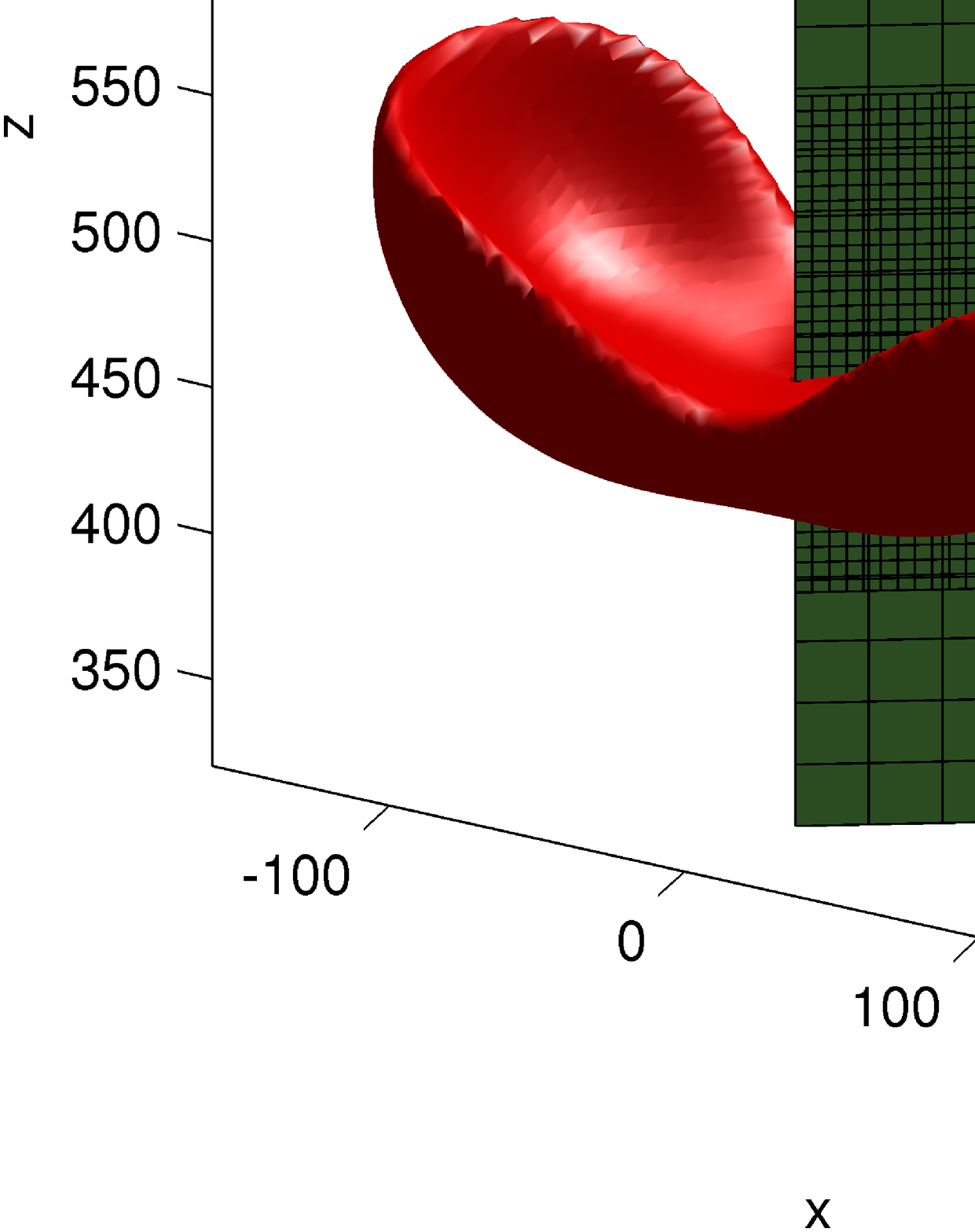}
\includegraphics[width=.45\textwidth]{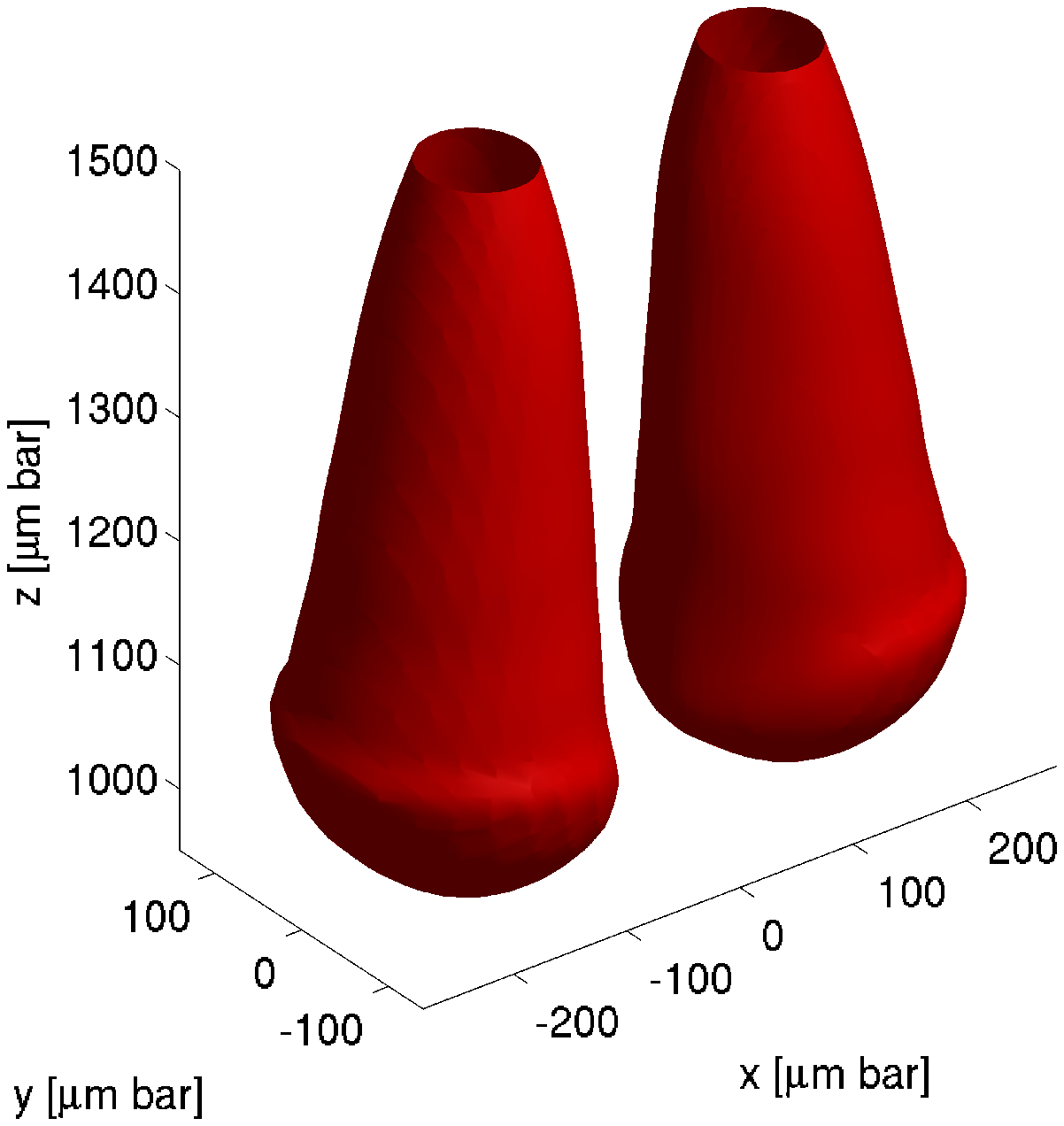}
\caption{\label{twostreamers}  
  \textbf{Left:} Scheme of refinement for the 3d simulation approach described     
  in the text.  The red surfaces surrounds the streamer head, defined  
  by its charge density.  The slices represent the numerical resolution of the density equations: they are equi-spaced in the angular direction but in the $r,z$ plane they contain a scheme of nested grids, common for all slices.
\textbf{Right:} Surfaces of constant electron density
  of two interacting negative streamers in nitrogen at atmospheric
  pressure propagating downwards in a
  homogeneous electric field of 80 kV/cm.  Reproduced 
  from \cite{Luque2008/PhRvL}.
}
\end{figure}

Another approach to the problem of streamer interaction was presented
in \cite{Luque2008/PhRvL}, where the method of 3d calculations
explained above was used to study the interaction between two negative
streamer heads propagating in parallel in a homogeneous electric
field (see Fig.~\ref{twostreamers}).

In general, for a given separation between the
streamer heads,  the outcome of the interaction depends on the gas composition.  In cases where 
photo-ionization is absent such as in pure nitrogen
(as far as it can be realized in the laboratory \cite{Nijdam2010/JPD}), the electrostatic 
repulsion between the charges in the
streamer heads drives them appart, bending the streamer channels
outwards.  However, for gas compositions where
photo-ionization plays an important role, such as air, the ionization in the volume
between the two heads counteracts the electrostatic repulsion between
them, driving the two streamers to merge.  In general the
outcome of the streamer interaction would therefore depend on the
separation between them and the balance of these two counteracting
processes.

\subsection{Branching}
\label{branching}
The branching of a streamer channel is perhaps the most 
challenging phenomenon for
numerical simulations.  The main reason is that branching is caused by a
dynamical instability at the streamer front, where small
irregularities in the very thin space-charge layer grow and eventually
form new streamers.  Therefore simulations of the branching process
demand a high numerical accuracy to rule out numerical artifacts.
This difficulty is compounded by the fact that branching completely breaks the
cylindrical symmetry of a single streamer channel, thus requiring full
3D simulations for the complete understanding of the phenomenon.

In 2002 Array{\'a}s \emph{et al.} \cite{Arrayas2002/PhRvL} studied
the branching of a negative streamer in a
minimal model that did not include photo-ionization.  There,
cylindrical symmetry was imposed and the result was interpreted as an
upper bound for branching.  The resulting picture was one of branching
due to a Laplacian instability, in close analogy to other well studied
physical models such as diffusion-limited aggregation and viscous
fingering.  Following this result, the numerical scheme was improved
by Montijn et al. \cite{Montijn2006/JCoPh}, who showed that branching
time converged as the highest resolution of the numerical grid was
improved \cite{Montijn2006/PhRvE}, thus ruling out that branching was
due to numerical artifacts.

Although those results imposed a cylindrical symmetry on the streamer,
we can show now that they are robust when that restriction is removed.
We use the method of 3D simulations detailed in
\cite{Luque2008/PhRvL} to investigate the dynamics of a negative streamer
without photo-ionization under small perturbations that break its cylindrical symmetry.  We study here only a linear regime in which the solution of our
system of equations remains at all times close to a cylindrically symmetrical solution; in a realistic situation the deviations from this solution are likely larger and the streamer will sooner deviate significantly from the symmetrical solution. 

In our simulation a negative streamer propagates in nitrogen
(i.e. there is no photo-ionization nor electron losses via attachment) between parallel plates in a 14.4 mm gap with a homogeneous field of 50 kV/cm.
For the spatial discretization we use the nested-grids approach 
of Montijn et al. \cite{Montijn2006/JCoPh}, discussed above.
The highest resolution in $r$ and $z$ is $\Delta r = \Delta z = 2.5\,
\mathrm{\mu m}$

\begin{figure}
\includegraphics[width=.6\textwidth]{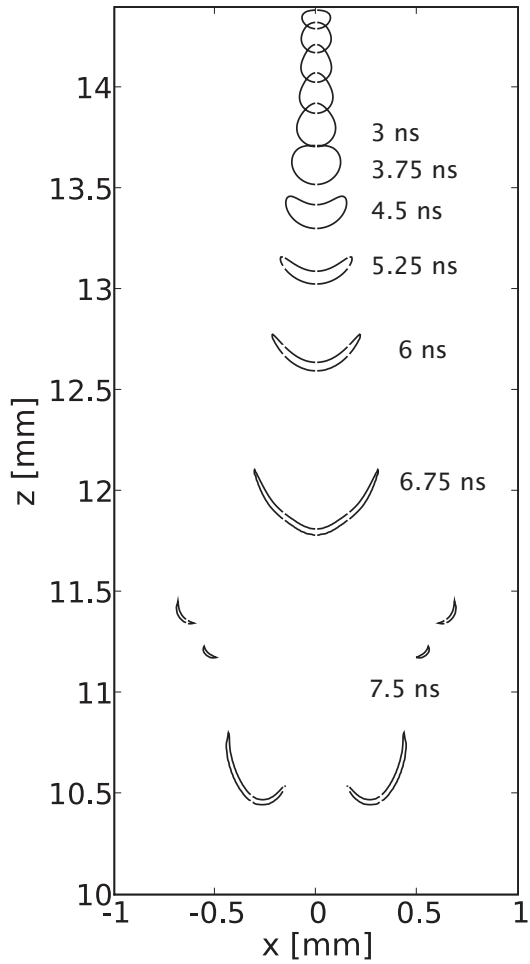} 
\caption{\label{evol} Evolution of an axi-symmetrical streamer head 
  represented by contours delimiting the region where charge density
  is larger than half of its maximum.  The contours are plotted 
  at regular timesteps of 0.75 ns.  Note that the equivalent figure
  for the evolution of a streamer with a slight perturbation of the
  initial symmetry looks almost identical to this one due to the
  smallness of the perturbation.}
\end{figure}

First we run a simulation with perfectly symmetrical initial
conditions consisting in a semispherical gaussian neutral ionization seed attached to the upper electrode.  The radius  of the seed is $73.6\units{\mu m}$ and its highest electron and ion density is $2.6\dexp{10}\units{cm^{-3}}$.
As seen in Fig.~\ref{evol}, the streamer branches between
6.75 and 7.5 ns.  In this context, ``branching'' means that
the streamer front loses its convexity.

To show that this branching is indeed associated with the breaking of
cylindrical symmetry observed in experiments, we run
a similar simulation where the initial condition deviates slightly
from cylindrical symmetry.  In this simulation we used cylindrical
coordinates with angular resolution $\Delta \theta = 2\pi / N$, $N=32$.

The asymmetrical initial state is constructed from 
the symmetrical initial electron densities $\bar{n_e}(r, z)$ by drawing
$\eta_1,\dots,\eta_{N-1}$ numbers from a normal distribution 
$\mathcal{N}(0, 1)$ and then setting
\begin{eqnarray}
  \net_0(r, z, t=0) & = & \bar{n_e}(r, z) \\
  \net_k(r, z, t=0) & = & \epsilon \eta_k \bar{n_e}(r, z)
                                \frac{r}{R_0} \;\;\;\;(k \neq 0),
\end{eqnarray}
where $\net_k(r, z, t)$ is the discrete Fourier transform of the
asymmetrical electron density, defined as in \parref{dft}:
\begin{equation}
  \tilde{n_e}_k(r, z) = \sum_{n=0}^{N-1} n_e(r, z, 2\pi n / N) 
    e^{-2\pi ikn/N}. 
\end{equation}
The strength of the perturbation to the
symmetrical state is $\epsilon$ and the factor
$r/R_0$ is introduced to keep $n_e(r, z, \theta)$ continuous at $r=0$.
The typical length $R_0$ is introduced to keep $\epsilon$
dimensionless.  Here we take $R_0=2.3\units{\mu m}$
(the inverse of the Townsend multiplication rate),
$\epsilon=10^{-8}$ but we also used $\epsilon=10^{-6}$ to check that
we were indeed in a linear regime and the growth rates of the
perturbation do not depend on $\epsilon$.
We chose to perturb only the initial electron density, keeping the
initial ion density symmetrical.

Note that this way of perturbing the electron density does not
guarantee that it will remain positive.  However, here we are
interested in very small perturbations; in this
limit the resulting density will almost certainly be positive
everywhere and we checked for that.

To track the small deviations from perfect symmetry observed in
the asymmetrical simulation we define
\begin{equation}
  W^{(u)}_k = 2\pi \Delta \theta \int_{-\infty}^{+\infty}dz\, \int_0^{\infty}dr\, r \,\tilde{u}_k(r, z),
\end{equation}
where $u$ stands for either the electron density $n_e$ or the charge density $n_i-n_e$. The $W^{(u)}_k$ give us an idea of the ``spectral content'' of electrons or charges in each mode.
Note that $W^{(n_e)}_0$ is the total number of electrons 
and $W^{(n_i - n_e)}_0$ is the total charge contained in the simulated volume.
In general the $W^{(u)}_k$ are $N$ complex numbers $k=0,\dots, N-1$ but
since they come from the DFT of a real quantity, they satisfy relations
analogous to \parref{dft-real}; for $N$ even they reduce to 
$(N - 2) / 2$ complex numbers and $2$ reals.  To analyze the spectral
content of the electron density we look at the amplitudes 
$|W^{(u)}_k| = \sqrt{W^{(u)}_k W^{(u)\star}_k} = \sqrt{W^{(u)}_k W^{(u)}_{N-k}}$ for $k=0,\dots,N / 2$. 

\begin{figure}
\includegraphics[width=.5\textwidth]{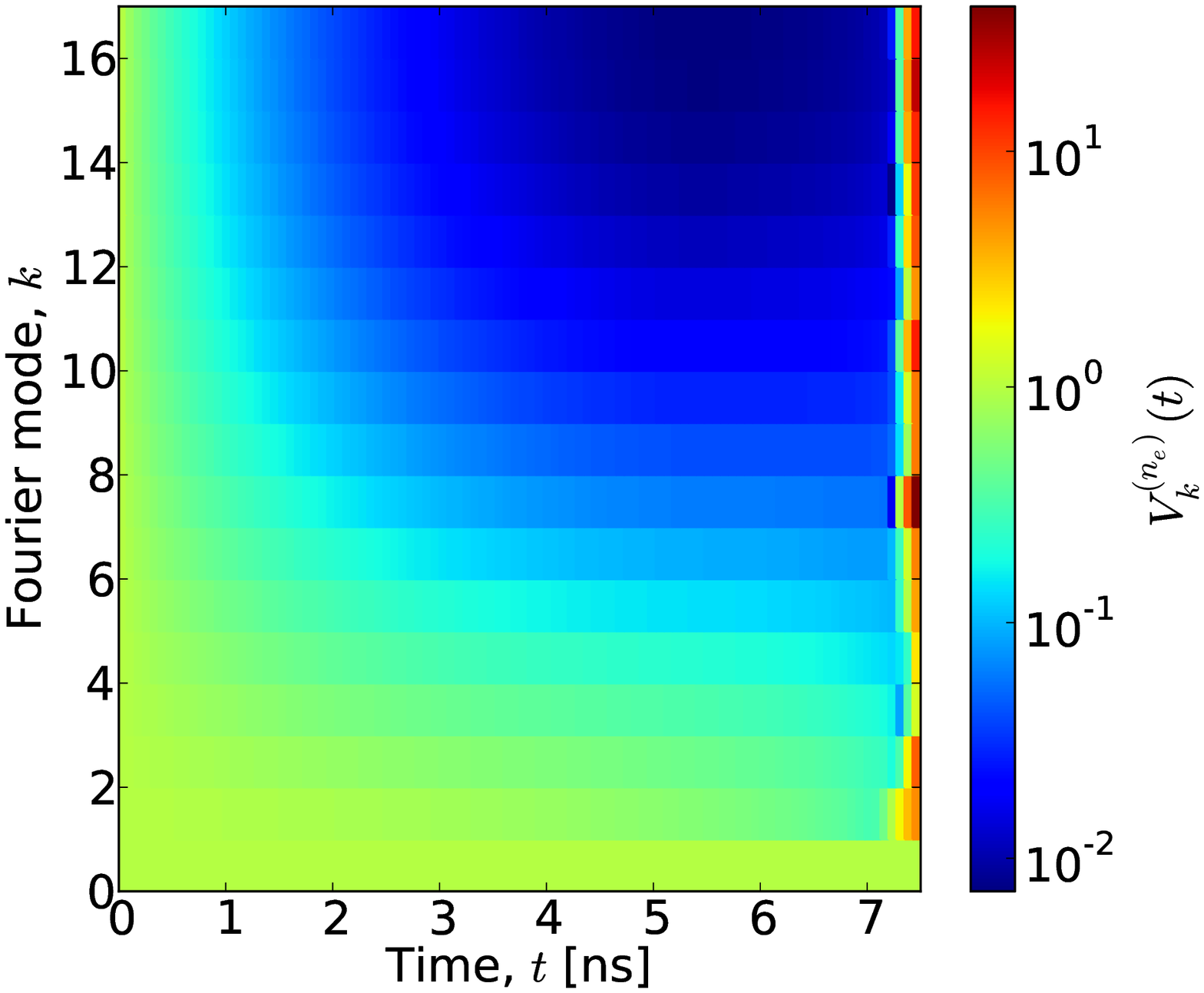} 
\includegraphics[width=.5\textwidth]{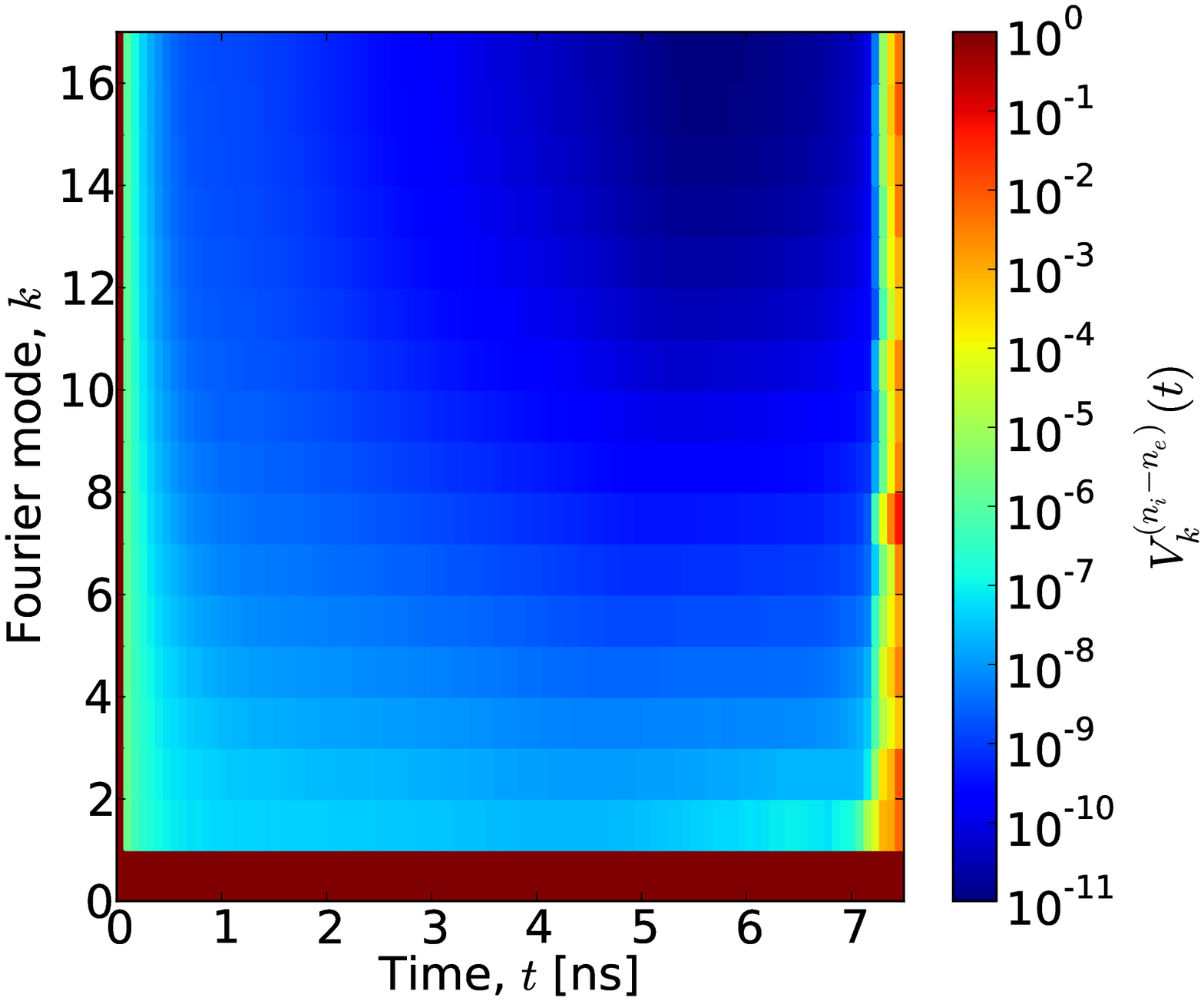} 
\caption{\label{vk} Time dependence of the normalized Fourier components of the electron density [$V^{(n_e)}_k(t)$; see text for a definition] (left) and
the space charge density [$V^{(n_i-n_e)}_k(t)$] (right) in a simulation with a slightly asymmetrical initial condition.  Note that the $V^{(u)}_k(t)$ with $k > 0$ decay smoothly up to $t = t_B \approx 7\units{ns}$. This
means that the $k=0$ component of the electron density (i.e. the total number of electrons) is growing faster than any other Fourier component; we interpret
this as the streamer becoming closer to cylindrical symmetry.  At $t = t_B$ a change in behavior is clear; after this points several $V^{(u)}_k(t)$ with $k > 0$ grow faster than the $k=0$ mode and the cylindrical symmetry is gradually broken.}
\end{figure}

Still, plotting $|W^{(u)}_k(t)|$ is usually not the best way to visualize the 
evolution of the
streamer.  First, the number of electrons and the total charge changes in time; one should
compare each mode with the corresponding total, $|W^{(u)}_0|$.  Second, because
we give random initial values to each $|W^{(u)}_k|$, it is also better to normalize
them according to their initial value.  Hence we define
\begin{equation}
  V^{(u)}_k(t) = \frac{|W^{(u)}_k(t)||W^{(u)}_0(0)|}{|W^{(u)}_0(t)||W^{(u)}_k(0)|},
  \label{Vk}
\end{equation}
which satisfies $V^{(u)}_0(t)=1$ for all times, and 
$V^{(u)}_k(0) = 1$ for all $k$. 
The evolution of the $V^{(u)}_k(t)$ is shown in Fig.~\ref{vk}.

\begin{figure}
\includegraphics[width=.85\textwidth]{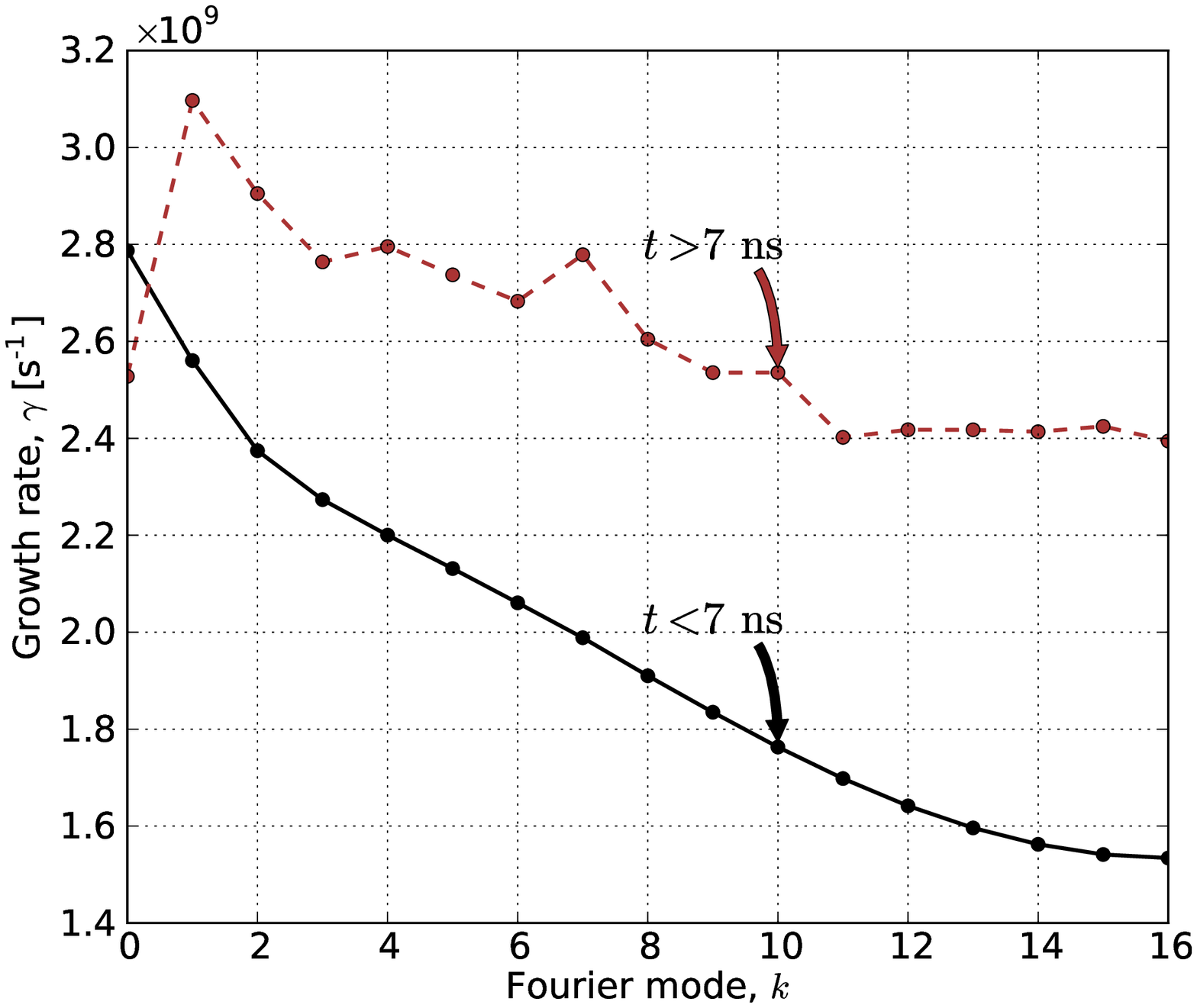} 
\caption{\label{rates} Growth rate of the unnormalized Fourier 
  components of the space charge density [$W^{(n_i-n_e)}_k$].
  We plot separately the growth rates before and after the
  branching time $t_B \approx 7\units{ns}$.  Note that for $t > t_B$ several 
  Fourier modes grow faster than the $k_0$ mode, indicating that the shape
  of the streamer deviates from cylindrical symmetry.
}
\end{figure}

The evolution of the axial perturbation clearly shows a change of
regime at $t=t_B$ with $t_B \approx 7 \,\mathrm{ns}$, coinciding with the onset of streamer branching.  For $t < t_B$ the evolution of the Fourier modes is
smooth and all of them decay with respect to the $k=0$ mode ($dV^{(n_e)}_k/dt
< 0$); during this phase the streamer
is becoming more symmetrical.  On the other hand for $t > t_B$ some
modes $k \ge 1$ grow with respect to the $k=0$ mode.  This is also
shown in Fig.~\ref{rates}, where we represented the growth rates
$\gamma$ of the unnormalized components $W^{(n_i-n_e)}_k$ of the space charge density; for each $k$, $\gamma$ is obtained by fitting $W^{(u)}_k(t)$ to an exponential $A e^{\gamma t}$ for $t < t_B$ and $t > t_B$.

These results show that ``branching'' interpreted as non-convexity of
an axi-symmetric solution is indeed related to faster growth of the
axial non-symmetric modes relative to the $k=0$ mode and hence to 
``branching'' in the more conventional sense of bifurcation of a single channel.

We note that there are two possible reasons for this outcome:
(a) that the streamer reaches a state of general dynamical instability that simultaneously affects both the cylindrically symmetrical and the asymmetrical modes and (b) that a non-convex, cylindrically symmetric streamer head is unstable against perturbations of its symmetry; this is expected if the electric field has a significant off-axis peak.  These two processes are compatible and likely both play a role.

\subsection{Streamers in inhomogeneous media: sprite streamers}
Up to now, we have only considered streamers propagating in
homogeneous media.  Streamers models have been extended to inhomogeneous gas densities and compositions in order to study the interaction between a streamer and gas bubbles \cite{Babaeva2008/ITPS} and the propagation of a streamer inside a plasma jet \cite{Naidis2010/JPhD}. 
Here we consider another important case where streamers
propagate in an inhomogeneous gas, namely sprites \cite{Franz1990/Sci}.  These are
transient discharges in the upper atmosphere above active
thunderclouds, where air density varies with altitude.  Sprites span altitudes from about 50 km up to 90 km,
are tens of kilometers wide and usually last for some milliseconds.
Often they are composed of two separate regions: an upper, diffuse one
usually extending from about 80 to 90 km altitude and a lower,
filamentary region composed by hundreds of streamers 
\cite{Raizer1998/JPhD, Pasko1998b/GeoRL}
(sometimes called ``sprite tendrils''). 
Pasko has recently reviewed models of single sprite streamers propagating in homogeneous air density \cite{Pasko2010/JGRA}.

A single streamer in the filamentary region can extend for
tens of kilometers in altitude.  In this range the density of air
varies considerably (roughly by a factor of 2 every 5 km).  Typical streamer 
lengths approximately scale with density \cite{Ebert2010/JGRA} 
so the density variation introduces another length 
scale on top of those discussed above.  The first numerical models to
include the variation of air density in sprite streamers were
published recently \cite{Luque2009/NatGe, Luque2010/GeoRL}.

\begin{figure}
\includegraphics[width=.6\textwidth]{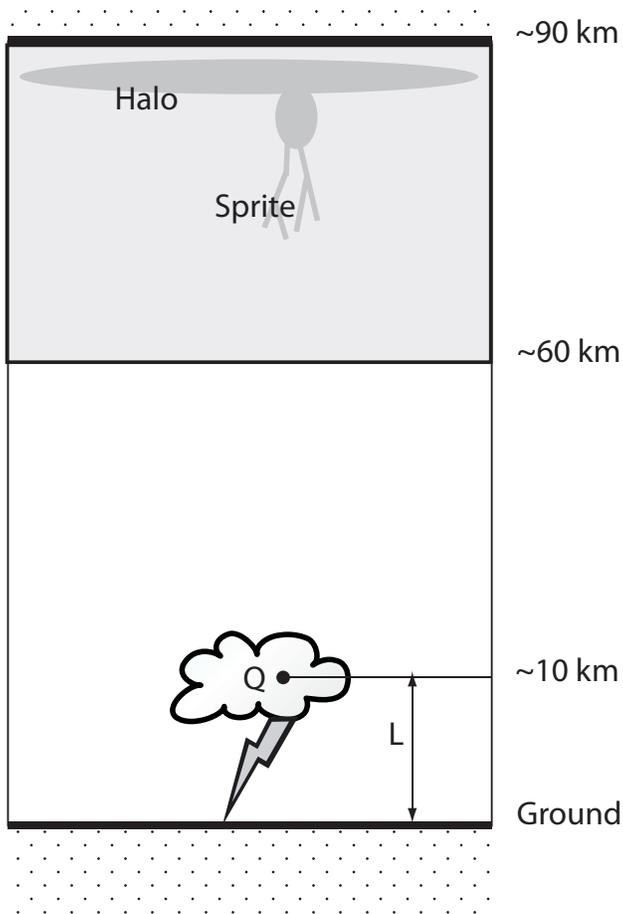} 
\caption{\label{thunder_scheme} Scheme of the geometry of sprite streamer 
  simulations.   First the field of the cloud charge with boundary conditions on ground and ionosphere is solved analytically.  The the Poisson equation for the sprite charges is solved in the complete domain
  between ground and ionosphere; the diffusion-drift-reaction
  equations are solved only in the upper, shaded volume.}
\end{figure}
 
The geometry of sprite simulations is depicted schematically in 
Fig.~\ref{thunder_scheme}.  The electrical field is created when a
thunderstroke at an altitude $L \approx 10\, \mathrm{km}$ moves a 
large charge to
the ground.  Most sprites are generated after positive cloud-to-ground
lightning strokes; these strokes leave a negative charge $Q$ in the cloud.
In the quasi-electrostatic approximation \cite{Pasko1997/JGR}
$Q$ is taken to change so slowly in time that the electromagnetic
radiative terms can be neglected (i.e. the speed of light is infinite).  The upper and lower boundaries of the resulting
electrostatic problem are, respectively, the Earth's surface and a
horizontal layer in the ionosphere at about 85-90 km altitude, 
where the Maxwell relaxation time due to the atmospheric conductivity 
is much shorter than the characteristic times involved in the discharge.

In principle, the charge $Q$ can be included in the simulation domain and treated as a prescribed source term in the Poisson equation.  However, the distance between this charge and the upper boundary of the simulation is about 80 km and in order to minimize the effects of the artificial lateral boundaries the domain would have to be extended to a radius much larger than these 80 km.  To avoid this overextension, in \cite{Luque2009/NatGe} the field created by $Q$ was calculated semi-analytically as the field created by a charge at a distance $L$ from the lower of two parallel electrodes.  This is expressed as a sum of the fields created by an infinite number of image charges; this sum is truncated after 4 terms (2 on each side).  To this one adds the self-consistent field of the sprite, calculated by solving the Poisson equation with homogeneous Dirichlet ($\phi=0$) boundary conditions in the upper and lower electrodes.  The horizontal width of the domain must be larger than the lateral extension of the space charges to minimize the effects of artificial boundaries but it can still be smaller that the distance between $Q$ and the sprite.

Once the field is calculated, the diffusion-drift-reaction equations
must only be solved in a smaller volume (see 
Fig.\ref{thunder_scheme}).  In \cite{Luque2009/NatGe} the volume was
a cylinder from 55 to 85 km in altitude, with a radius of 20 km.
 
A notable difference between streamers in the laboratory and sprite
streamers is that in a laboratory they are almost always initiated
from a sharp metallic electrode that is clearly not present in the upper
atmosphere.  Raizer \cite{Raizer1998/JPhD} proposed that streamers
emerge from ``plasma patches'' created by electro-magnetic waves
\cite{Valdivia1997/GeoRL}.  Recently, Marshall et al. 
\cite{Marshall2010/JGRA/1} used a finite
difference, full electromagnetic code to evaluate the influence of
lightning-generated electromagnetic pulses (EMP) on the electron density in
the upper atmosphere, reporting large enhancements of up to a factor 4
for repeated EMP's.

\begin{figure}
\includegraphics[width=.85\textwidth]{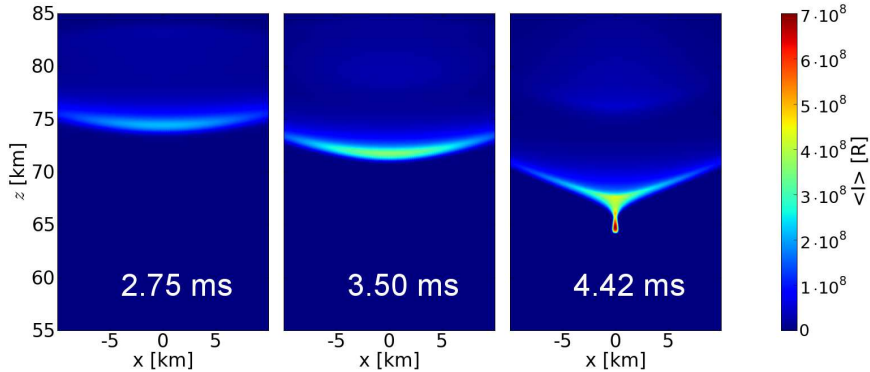} 
\caption{\label{sprite} Optical emissions emitted from a streamer
  emerging from the sharpened edge of a relaxation-ionization wave
  from the ionosphere.  This wave produces diffuse optical emissions
  observed as a halo. (Reproduced from \cite{Luque2009/NatGe})}
\end{figure}

However, an isolated patch would create a two-headed streamer propagating 
simultaneously up and down \cite{Liu2004/JGRA/1}.  In observations, however, sprite streamers
always start propagating downwards 
\cite{Cummer2006/GeoRL,Stenbaek-Nielsen2008/JPhD}.  In 
\cite{Luque2009/NatGe} it was 
shown how a single, downwards-propagating sprite streamer emerges from the destabilization 
of a sharp screening-ionization wave from the ionosphere;
figure~\ref{sprite} shows the optical emissions resulting from that process.
The initial condition amounted here to air density exponentially decreasing with altitude and electron density increasing with altitude as in fair weather night-time.

A relevant recent result related to sprite streamers concerns the
trailing optical emissions emanating from the streamer body at a
certain distance from the head.  Although initially this luminosity
was attributed to chemical processes in the wake of the streamer head
\cite{Kanmae2007/GeoRL, Sentman2008/JGRD},
two recent papers \cite{Liu2010/GeoRL, Luque2010/GeoRL} used streamer 
density models to show that it originates from a re-enhancement of the
electric field.  This ocurs if there is a substantial electrical current in the streamer body, generated because the net charge in the streamer head increases during propagation.  In \cite{Luque2010/GeoRL} the electric field in this re-enhancement approached $E\approx E_k$; this appears to be dynamically stable in a current-carrying state.
 
\section{Outlook}
As with any scientific endeavour, it is hard to predict the future of streamer density models.  The ultimate aim of simulating and understanding a fully branched tree of streamers strictly within the model described in this paper
looks unrealistic in the short and medium term.  

On the other hand, gradual improvements to density models will surely come in the short run. There are two obvious directions to extend the physical models.  In the ``microscopic direction'' the modeling of the microscopic phenomena is improved; this includes e.g. a more accurate calculation of the photo-ionization sources, additional chemical processes and the inclusion of stochasticity, converging towards detailed particle and hybrid models. 
The ``macroscopic direction,'' extends the simulation domain, possibly to include many interacting streamers and also incorporate in an improved fashion the interaction between the streamer and its surroundings.
Although in principle they are compatible, both directions demand a higher computational it is unlikely that they will be combined in a practicable unified model.  Even if approached separately, they will likely require faster computers and numerical algorithms. 

However, it may not be necessary to describe the full range of physics, from the microscopic to the macroscopic level, inside a single model.  A more promising approach is to develop a hierarchy of models of decreasing microscopic detail.  Each model would rely on simplifications and assumptions extracted from models down in the hierarchy or from direct experimental observations. 
Density models play a pivotal role in this hierarchy, sitting between particle models and higher-level descriptions of a streamer tree such as the dielectric breakdown model of Niemeyer et al. \cite{Niemeyer1984/PhRvL,Pasko2000/GeoRL}, phenomenological branching models \cite{Akyuz2003/JElec} and moving boundary models \cite{Lozansky1973/JPhD,Meulenbroek2004/PhRvE, Brau2008/PhRvE/1,Brau2008/PhRvE,Brau2009/PhRvE}.

\section*{Acknowledgements}
We thank C. Li for useful comments on this paper.  AL was supported by
the Spanish Ministry of Science and Innovation, MICINN under project AYA2009-14027-C05-02.

\newcommand{\jgr}{J. Geophys. Res.\xspace}
\newcommand{\grl}{Geophys. Res. Lett.\xspace}
\newcommand{\pra}{Phys. Rev. A\xspace}
\newcommand{\prb}{Phys. Rev. B \xspace}
\newcommand{\prc}{Phys. Rev. C \xspace}
\newcommand{\prd}{Phys. Rev. D \xspace}
\newcommand{\pre}{Phys. Rev. E \xspace}

\bibliographystyle{elsarticle-num}
\bibliography{Everything}

\begin{thebibliography}{10}
\expandafter\ifx\csname url\endcsname\relax
  \def\url#1{\texttt{#1}}\fi
\expandafter\ifx\csname urlprefix\endcsname\relax\def\urlprefix{URL }\fi
\expandafter\ifx\csname href\endcsname\relax
  \def\href#1#2{#2} \def\path#1{#1}\fi

\bibitem{Raether1939/ZPhy}
H.~{Raether}, {Die Entwicklung der Elektronenlawine in den Funkenkanal},
  Zeitschrift fur Physik 112 (1939) 464.
\newblock \href {http://dx.doi.org/10.1007/BF01340229}
  {\path{doi:10.1007/BF01340229}}.

\bibitem{Raizer1991/book}
Y.~P. Raizer, {Gas Discharge Physics}, Springer-Verlag, Berlin, Germany, 1991.

\bibitem{Ebert2006/PSST}
U.~{Ebert}, C.~{Montijn}, T.~M.~P. {Briels}, W.~{Hundsdorfer},
  B.~{Meulenbroek}, A.~{Rocco}, E.~M. {van Veldhuizen}, {The multiscale nature
  of streamers}, Plasma Sour. Sci. Technol. 15 (2006) 118.
\newblock \href {http://arxiv.org/abs/physics/0604023}
  {\path{arXiv:physics/0604023}}, \href
  {http://dx.doi.org/10.1088/0963-0252/15/2/S14}
  {\path{doi:10.1088/0963-0252/15/2/S14}}.

\bibitem{Ebert2008/JPhD}
U.~{Ebert}, D.~D. {Sentman}, {EDITORIAL REVIEW: Streamers, sprites, leaders,
  lightning: from micro- to macroscales}, J. Phys. D 41~(23) (2008) 230301.
\newblock \href {http://arxiv.org/abs/0811.2075} {\path{arXiv:0811.2075}},
  \href {http://dx.doi.org/10.1088/0022-3727/41/23/230301}
  {\path{doi:10.1088/0022-3727/41/23/230301}}.

\bibitem{Li2011/JCP}
C.~{Li}, U.~{Ebert}, W.~{Hundsdorfer}, {Spatially hybrid computations for
  streamer discharges: II. fully 3D computations}, J. Comput. Phys., submitted
  to the same special issue as the present article.

\bibitem{Lozansky1973/JPhD}
E.~D. {Lozansky}, O.~B. {Firsov}, {Theory of the initial stage of streamer
  propagation}, J. Phys. D 6 (1973) 976.
\newblock \href {http://dx.doi.org/10.1088/0022-3727/6/8/310}
  {\path{doi:10.1088/0022-3727/6/8/310}}.

\bibitem{Meulenbroek2004/PhRvE}
B.~{Meulenbroek}, A.~{Rocco}, U.~{Ebert}, {Streamer branching rationalized by
  conformal mapping techniques}, \pre 69~(6) (2004) 067402.
\newblock \href {http://arxiv.org/abs/physics/0305112}
  {\path{arXiv:physics/0305112}}, \href
  {http://dx.doi.org/10.1103/PhysRevE.69.067402}
  {\path{doi:10.1103/PhysRevE.69.067402}}.

\bibitem{Brau2008/PhRvE/1}
F.~{Brau}, A.~{Luque}, B.~{Meulenbroek}, U.~{Ebert}, L.~{Sch{\"a}fer},
  {Construction and test of a moving boundary model for negative streamer
  discharges}, \pre 77~(2) (2008) 026219.
\newblock \href {http://arxiv.org/abs/0707.1402} {\path{arXiv:0707.1402}},
  \href {http://dx.doi.org/10.1103/PhysRevE.77.026219}
  {\path{doi:10.1103/PhysRevE.77.026219}}.

\bibitem{Brau2008/PhRvE}
F.~{Brau}, B.~{Davidovitch}, U.~{Ebert}, {Moving boundary approximation for
  curved streamer ionization fronts: Solvability analysis}, \pre 78~(5) (2008)
  056212.
\newblock \href {http://arxiv.org/abs/0807.4614} {\path{arXiv:0807.4614}},
  \href {http://dx.doi.org/10.1103/PhysRevE.78.056212}
  {\path{doi:10.1103/PhysRevE.78.056212}}.

\bibitem{Brau2009/PhRvE}
F.~{Brau}, A.~{Luque}, B.~{Davidovitch}, U.~{Ebert}, {Moving-boundary
  approximation for curved streamer ionization fronts: Numerical tests}, \pre
  79~(6) (2009) 066211.
\newblock \href {http://arxiv.org/abs/0901.1916} {\path{arXiv:0901.1916}},
  \href {http://dx.doi.org/10.1103/PhysRevE.79.066211}
  {\path{doi:10.1103/PhysRevE.79.066211}}.

\bibitem{Li2010/JCoPh}
C.~{Li}, U.~{Ebert}, W.~{Hundsdorfer}, {Spatially hybrid computations for
  streamer discharges with generic features of pulled fronts: I. Planar
  fronts}, J. Comput. Phys. 229 (2010) 200.
\newblock \href {http://arxiv.org/abs/0904.2968} {\path{arXiv:0904.2968}},
  \href {http://dx.doi.org/10.1016/j.jcp.2009.09.027}
  {\path{doi:10.1016/j.jcp.2009.09.027}}.

\bibitem{Li2007/JAP}
C.~{Li}, W.~J.~M. {Brok}, U.~{Ebert}, J.~J.~A.~M. {van der Mullen}, {Deviations
  from the local field approximation in negative streamer heads}, J. Appl.
  Phys. 101~(12) (2007) 123305.
\newblock \href {http://arxiv.org/abs/physics/0702129}
  {\path{arXiv:physics/0702129}}, \href {http://dx.doi.org/10.1063/1.2748673}
  {\path{doi:10.1063/1.2748673}}.

\bibitem{Li2009/JPhD}
C.~{Li}, U.~{Ebert}, W.~{Hundsdorfer}, {FAST TRACK COMMUNICATION: 3D hybrid
  computations for streamer discharges and production of runaway electrons}, J.
  Phys. D 42~(20) (2009) 202003.
\newblock \href {http://arxiv.org/abs/0907.0555} {\path{arXiv:0907.0555}},
  \href {http://dx.doi.org/10.1088/0022-3727/42/20/202003}
  {\path{doi:10.1088/0022-3727/42/20/202003}}.

\bibitem{Chanrion2010/JGR}
O.~{Chanrion}, T.~{Neubert}, Production of runaway electrons by negative
  streamer discharges, J. Geophys. Res. - Space Phys. 115.
\newblock \href {http://dx.doi.org/???} {\path{doi:???}}

\bibitem{Phelps1987/PhRvA}
A.~V. {Phelps}, B.~M. {Jelenkovich}, L.~C. {Pitchford}, {Simplified models of
  electron excitation and ionization at very high E/n}, \pra 36 (1987) 5327.
\newblock \href {http://dx.doi.org/10.1103/PhysRevA.36.5327}
  {\path{doi:10.1103/PhysRevA.36.5327}}.

\bibitem{Moss2006/JGRA}
G.~D. {Moss}, V.~P. {Pasko}, N.~{Liu}, G.~{Veronis}, {Monte Carlo model for
  analysis of thermal runaway electrons in streamer tips in transient luminous
  events and streamer zones of lightning leaders}, J. Geophys. Res. (Space
  Phys) 111 (2006) 2307.
\newblock \href {http://dx.doi.org/10.1029/2005JA011350}
  {\path{doi:10.1029/2005JA011350}}.

\bibitem{Vrhovac2001/JAP}
S.~B. {Vrhovac}, V.~D. {Stojanovi{\'c}}, B.~M. {Jelenkovi{\'c}}, Z.~L.
  {Petrovi{\'c}}, {Energy distributions of electrons in a low-current
  self-sustained nitrogen discharge}, J. Appl. Phys. 90 (2001) 5871.
\newblock \href {http://dx.doi.org/10.1063/1.1415364}
  {\path{doi:10.1063/1.1415364}}.

\bibitem{Ebert2010/JGRA}
U.~{Ebert}, S.~{Nijdam}, C.~{Li}, A.~{Luque}, T.~{Briels}, E.~{van Veldhuizen},
  {Review of recent results on streamer discharges and discussion of their
  relevance for sprites and lightning}, J. Geophys. Res. (Space Phys) 115
  (2010) 0.
\newblock \href {http://arxiv.org/abs/1002.0070} {\path{arXiv:1002.0070}},
  \href {http://dx.doi.org/10.1029/2009JA014867}
  {\path{doi:10.1029/2009JA014867}}.

\bibitem{Pasko1998b/GeoRL}
V.~P. {Pasko}, U.~S. {Inan}, T.~F. {Bell}, {Spatial structure of sprites},
  Geophys. Res. Lett. 25 (1998) 2123.
\newblock \href {http://dx.doi.org/10.1029/98GL02123 CORRECT?}
  {\path{doi:10.1029/98GL02123 CORRECT?}}

\bibitem{Wormeester2010/arxiv}
G.~{Wormeester}, S.~{Pancheshnyi}, A.~{Luque}, S.~{Nijdam}, U.~{Ebert},
  {Probing photo-ionization: simulations of positive streamers in varying N2:O2
  mixtures}\href {http://arxiv.org/abs/1008.3309} {\path{arXiv:1008.3309}}.

\bibitem{Nijdam2010/JPD}
S.~Nijdam, F.~M. J. H.~v. de~Wetering, R.~Blanc, E.~M. van Veldhuizen,
  U.~Ebert, {Probing photo-ionization: experiments on positive streamers in
  pure gases and mixtures}, {J. Phys. D: Appl. Phys.} {43}~({14}).
\newblock \href {http://dx.doi.org/{10.1088/0022-3727/43/14/145204}}
  {\path{doi:{10.1088/0022-3727/43/14/145204}}}.

\bibitem{Montijn2006/JCoPh}
C.~{Montijn}, W.~{Hundsdorfer}, U.~{Ebert}, {An adaptive grid refinement
  strategy for the simulation of negative streamers}, J. Comput. Phys. 219
  (2006) 801.
\newblock \href {http://arxiv.org/abs/physics/0603070}
  {\path{arXiv:physics/0603070}}, \href
  {http://dx.doi.org/10.1016/j.jcp.2006.04.017}
  {\path{doi:10.1016/j.jcp.2006.04.017}}.

\bibitem{Chanrion2008/JCoPh}
O.~{Chanrion}, T.~{Neubert}, {A PIC-MCC code for simulation of streamer
  propagation in air}, J. Comput. Phys. 227 (2008) 7222.
\newblock \href {http://dx.doi.org/10.1016/j.jcp.2008.04.016}
  {\path{doi:10.1016/j.jcp.2008.04.016}}.

\bibitem{Soria-Hoyo2009/JCoPh}
C.~{Soria-Hoyo}, F.~{Pontiga}, A.~{Castellanos}, {A PIC based procedure for the
  integration of multiple time scale problems in gas discharge physics}, J.
  Comput. Phys. 228 (2009) 1017.
\newblock \href {http://dx.doi.org/10.1016/j.jcp.2008.10.007}
  {\path{doi:10.1016/j.jcp.2008.10.007}}.

\bibitem{Pancheshnyi2005/PSST}
S.~{Pancheshnyi}, {Role of electronegative gas admixtures in streamer start,
  propagation and branching phenomena}, Plasma Sour. Sci. Technol. 14 (2005)
  645.
\newblock \href {http://dx.doi.org/10.1088/0963-0252/14/4/002}
  {\path{doi:10.1088/0963-0252/14/4/002}}.

\bibitem{Zhelezniak1982/TepVT}
M.~B. {Zhelezniak}, A.~K. {Mnatsakanian}, S.~V. {Sizykh}, {Photoionization of
  nitrogen and oxygen mixtures by radiation from a gas discharge}, Teplofizika
  Vysokikh Temperatur 20 (1982) 423.

\bibitem{Dubrovin2010/JGR}
D.~Dubrovin, S.~Nijdam, E.~M. van Veldhuizen, U.~Ebert, Y.~Yair, C.~Price,
  {Sprite discharges on Venus and Jupiter-like planets: A laboratory
  investigation}, {J. Geophys. Res - Space Phys.} {115}.
\newblock \href {http://dx.doi.org/{10.1029/2009JA014851}}
  {\path{doi:{10.1029/2009JA014851}}}.

\bibitem{Phelps1985/PhRvA}
A.~V. {Phelps}, L.~C. {Pitchford}, {Anisotropic scattering of electrons by N2
  and its effect on electron transport}, \pra 31 (1985) 2932.
\newblock \href {http://dx.doi.org/10.1103/PhysRevA.31.2932}
  {\path{doi:10.1103/PhysRevA.31.2932}}.

\bibitem{Hagelaar2005/PSST}
G.~J.~M. {Hagelaar}, L.~C. {Pitchford}, {Solving the Boltzmann equation to
  obtain electron transport coefficients and rate coefficients for fluid
  models}, Plasma Sour. Sci. Technol. 14 (2005) 722.
\newblock \href {http://dx.doi.org/10.1088/0963-0252/14/4/011}
  {\path{doi:10.1088/0963-0252/14/4/011}}.

\bibitem{Dujko2010/POAB}
S.~Dujko, U.~Ebert, R.~White, Z.~L. Petrovic, Electron transport data in
  n$_2$-o$_2$ streamer plasma discharges, {Publ. Astron. Obs. Belgrade} {89}
  ({2010}) 71--74.

\bibitem{Dhali1985/PhRvA}
S.~K. {Dhali}, P.~F. {Williams}, {Numerical simulation of streamer propagation
  in nitrogen at atmospheric pressure}, \pra 31 (1985) 1219.
\newblock \href {http://dx.doi.org/10.1103/PhysRevA.31.1219}
  {\path{doi:10.1103/PhysRevA.31.1219}}.

\bibitem{Dhali1987/JAP}
S.~K. {Dhali}, P.~F. {Williams}, {Two-dimensional studies of streamers in
  gases}, J. Appl. Phys. 62 (1987) 4696.
\newblock \href {http://dx.doi.org/10.1063/1.339020}
  {\path{doi:10.1063/1.339020}}.

\bibitem{Wu1988/PhRvA}
C.~{Wu}, E.~E. {Kunhardt}, {Formation and propagation of streamers in N$_{2}$
  and N$_{2}$-SF$_{6}$ mixtures}, \pra 37 (1988) 4396.
\newblock \href {http://dx.doi.org/10.1103/PhysRevA.37.4396}
  {\path{doi:10.1103/PhysRevA.37.4396}}.

\bibitem{Vitello1994/PhRvE}
P.~A. {Vitello}, B.~M. {Penetrante}, J.~N. {Bardsley}, {Simulation of
  negative-streamer dynamics in nitrogen}, \pre 49 (1994) 5574.
\newblock \href {http://dx.doi.org/10.1103/PhysRevE.49.5574}
  {\path{doi:10.1103/PhysRevE.49.5574}}.

\bibitem{Zalesak1979/JCoPh}
S.~T. {Zalesak}, {Fully multidimensional flux-corrected transport algorithms
  for fluids}, J. Comput. Phys. 31 (1979) 335.
\newblock \href {http://dx.doi.org/10.1016/0021-9991(79)90051-2}
  {\path{doi:10.1016/0021-9991(79)90051-2}}.

\bibitem{Kulikovsky1995/JCoPh}
A.~{Kulikovsky}, {A More Accurate Scharfetter-Gummel Algorithm of Electron
  Transport for Semiconductor and Gas Discharge Simulation}, J. Comput. Phys.
  119 (1995) 149.
\newblock \href {http://dx.doi.org/10.1006/jcph.1995.1123}
  {\path{doi:10.1006/jcph.1995.1123}}.

\bibitem{Liu2004/JGRA/1}
N.~{Liu}, V.~P. {Pasko}, {Effects of photoionization on propagation and
  branching of positive and negative streamers in sprites}, J. Geophys. Res.
  (Space Phys) 109 (2004) 4301.
\newblock \href {http://dx.doi.org/10.1029/2003JA010064}
  {\path{doi:10.1029/2003JA010064}}.

\bibitem{Bourdon2007/PSST}
A.~{Bourdon}, V.~P. {Pasko}, N.~Y. {Liu}, S.~{C{\'e}lestin}, P.~{S{\'e}gur},
  E.~{Marode}, {Efficient models for photoionization produced by non-thermal
  gas discharges in air based on radiative transfer and the Helmholtz
  equations}, Plasma Sour. Sci. Technol. 16 (2007) 656.
\newblock \href {http://dx.doi.org/10.1088/0963-0252/16/3/026}
  {\path{doi:10.1088/0963-0252/16/3/026}}.

\bibitem{Pancheshnyi2003/JPhD}
S.~V. {Pancheshnyi}, A.~Y. {Starikovskii}, {Two-dimensional numerical modelling
  of the cathode-directed streamer development in a long gap at high voltage},
  J. Phys. D 36 (2003) 2683.
\newblock \href {http://dx.doi.org/10.1088/0022-3727/36/21/014}
  {\path{doi:10.1088/0022-3727/36/21/014}}.

\bibitem{Koren1993/inbook}
B.~{Koren}, {A Robust Upwind Discretization Method for Advection, Diffusion and
  Source Terms}, in: {Vreugdenhil, C. B. and Koren, B} (Ed.), Numerical Methods
  for Advection-Diffusion Problems, 1993, pp. 117--137.

\bibitem{Kulikovsky2000/JPhD}
A.~A. {Kulikovsky}, {The role of photoionization in positive streamer
  dynamics}, J. Phys. D 33 (2000) 1514.
\newblock \href {http://dx.doi.org/10.1088/0022-3727/33/12/314}
  {\path{doi:10.1088/0022-3727/33/12/314}}.

\bibitem{Pancheshnyi2005/PhRvE}
S.~{Pancheshnyi}, M.~{Nudnova}, A.~{Starikovskii}, {Development of a
  cathode-directed streamer discharge in air at different pressures: Experiment
  and comparison with direct numerical simulation}, \pre 71~(1) (2005) 016407.
\newblock \href {http://dx.doi.org/10.1103/PhysRevE.71.016407}
  {\path{doi:10.1103/PhysRevE.71.016407}}.

\bibitem{Eichwald2008/JPhD}
O.~{Eichwald}, O.~{Ducasse}, D.~{Dubois}, A.~{Abahazem}, N.~{Merbahi},
  M.~{Benhenni}, M.~{Yousfi}, {Experimental analysis and modelling of positive
  streamer in air: towards an estimation of O and N radical production}, J.
  Phys. D 41~(23) (2008) 234002.
\newblock \href {http://dx.doi.org/10.1088/0022-3727/41/23/234002}
  {\path{doi:10.1088/0022-3727/41/23/234002}}.

\bibitem{Min2001/ITM}
W.~{Min}, H.~{Kim}, S.~{Lee}, S.~{Hahn}, {A study on the streamer simulation
  using adaptive mesh generation and FEM-FCT}, IEEE Transactions on Magnetics
  37 (2001) 3141.
\newblock \href {http://dx.doi.org/10.1109/20.952562}
  {\path{doi:10.1109/20.952562}}.

\bibitem{Nikandrov2008/ITPS}
D.~S. {Nikandrov}, R.~R. {Arslanbekov}, V.~I. {Kolobov}, {Streamer Simulations
  With Dynamically Adaptive Cartesian Mesh}, IEEE Trans. Plasma Sci. 36 (2008)
  932.
\newblock \href {http://dx.doi.org/10.1109/TPS.2008.924533}
  {\path{doi:10.1109/TPS.2008.924533}}.

\bibitem{Ebert2000/PhyD}
U.~{Ebert}, W.~{van Saarloos}, {Front propagation into unstable states:
  universal algebraic convergence towards uniformly translating pulled fronts},
  Physica D Nonlinear Phenomena 146 (2000) 1.
\newblock \href {http://arxiv.org/abs/cond-mat/0003181}
  {\path{arXiv:cond-mat/0003181}}, \href
  {http://dx.doi.org/10.1016/S0167-2789(00)00068-3}
  {\path{doi:10.1016/S0167-2789(00)00068-3}}.

\bibitem{Kulikovsky1997/JPhD/1}
A.~A. {Kulikovsky}, {Positive streamer between parallel plate electrodes in
  atmospheric pressure air}, J. Phys. D 30 (1997) 441.
\newblock \href {http://dx.doi.org/10.1088/0022-3727/30/3/017}
  {\path{doi:10.1088/0022-3727/30/3/017}}.

\bibitem{Pancheshnyi2001/JPhD/1}
S.~V. {Pancheshnyi}, S.~M. {Starikovskaia}, A.~Y. {Starikovskii}, {Role of
  photoionization processes in propagation of cathode-directed streamer}, J.
  Phys. D 34 (2001) 105.
\newblock \href {http://dx.doi.org/10.1088/0022-3727/34/1/317}
  {\path{doi:10.1088/0022-3727/34/1/317}}.

\bibitem{Wackers2005/JCoAM}
J.~{Wackers}, {A nested-grid direct Poisson solver for concentrated source
  terms}, J. Comput. App. Math. 180 (2005) 1.

\bibitem{Celestin2009/JPhD}
S.~{Celestin}, Z.~{Bonaventura}, B.~{Zeghondy}, A.~{Bourdon}, P.~{S{\'e}gur},
  {The use of the ghost fluid method for Poisson's equation to simulate
  streamer propagation in point-to-plane and point-to-point geometries}, J.
  Phys. D 42~(6) (2009) 065203.
\newblock \href {http://dx.doi.org/10.1088/0022-3727/42/6/065203}
  {\path{doi:10.1088/0022-3727/42/6/065203}}.

\bibitem{Kulikovsky1998/PhLA}
A.~A. {Kulikovsky}, {Three-dimensional simulation of a positive streamer in air
  near curved anode}, Phys. Lett. A 245 (1998) 445.
\newblock \href {http://dx.doi.org/10.1016/S0375-9601(98)00415-0}
  {\path{doi:10.1016/S0375-9601(98)00415-0}}.

\bibitem{Luque2008/PhRvL}
A.~{Luque}, U.~{Ebert}, W.~{Hundsdorfer}, {Interaction of Streamer Discharges
  in Air and Other Oxygen-Nitrogen Mixtures}, Phys. Rev. Lett. 101~(7) (2008)
  075005.
\newblock \href {http://arxiv.org/abs/0712.2774} {\path{arXiv:0712.2774}},
  \href {http://dx.doi.org/10.1103/PhysRevLett.101.075005}
  {\path{doi:10.1103/PhysRevLett.101.075005}}.

\bibitem{Babaeva1996/JPhD}
N.~Y. {Babaeva}, G.~V. {Naidis}, {Two-dimensional modelling of positive
  streamer dynamics in non-uniform electric fields in air}, J. Phys. D 29
  (1996) 2423.
\newblock \href {http://dx.doi.org/10.1088/0022-3727/29/9/029}
  {\path{doi:10.1088/0022-3727/29/9/029}}.

\bibitem{Luque2008/JPhD}
A.~{Luque}, V.~{Ratushnaya}, U.~{Ebert}, {Positive and negative streamers in
  ambient air: modelling evolution and velocities}, J. Phys. D 41~(23) (2008)
  234005.
\newblock \href {http://arxiv.org/abs/0804.3539} {\path{arXiv:0804.3539}},
  \href {http://dx.doi.org/10.1088/0022-3727/41/23/234005}
  {\path{doi:10.1088/0022-3727/41/23/234005}}.

\bibitem{Singer1974/IEEE}
H.~Singer, H.~Steinbigler, P.~Weiss, A charge simulation method for the
  calculation of high voltage fields, IEEE Trans. on Power Apparatus and
  Systems PAS-93~(5) (1974) 1660 --1668.
\newblock \href {http://dx.doi.org/10.1109/TPAS.1974.293898}
  {\path{doi:10.1109/TPAS.1974.293898}}.

\bibitem{Malik1989/IEEE}
N.~{Malik}, {A review of the charge simulation method and its applications},
  IEEE Trans. on Electrical Insulation 24 (1989) 3--20.

\bibitem{Li2009/PhDT}
C.~{Li}, \href{http://repository.tue.nl/640104}{{Joining particle and fluid
  aspects in streamer simulations}}, Ph.D. thesis, Technische Universiteit
  Eindhoven (2009).
\newline\urlprefix\url{http://repository.tue.nl/640104}

\bibitem{Serdyuk2001/JPhD}
Y.~V. {Serdyuk}, A.~{Larsson}, S.~M. {Gubanski}, M.~{Akyuz}, {The propagation
  of positive streamers in a weak and uniform background electric field}, J.
  Phys. D 34 (2001) 614.
\newblock \href {http://dx.doi.org/10.1088/0022-3727/34/4/323}
  {\path{doi:10.1088/0022-3727/34/4/323}}.

\bibitem{van_Dijk2009/JPhD}
J.~{van Dijk}, K.~{Peerenboom}, M.~{Jimenez}, D.~{Mihailova}, J.~{van der
  Mullen}, {REVIEW ARTICLE: The plasma modelling toolkit Plasimo}, J. Phys. D
  42~(19) (2009) 194012.
\newblock \href {http://dx.doi.org/10.1088/0022-3727/42/19/194012}
  {\path{doi:10.1088/0022-3727/42/19/194012}}.

\bibitem{Steinle1999/JPhD}
G.~{Steinle}, D.~{Neundorf}, W.~{Hiller}, M.~{Pietralla}, {Two-dimensional
  simulation of filaments in barrier discharges}, J. Phys. D 32 (1999) 1350.
\newblock \href {http://dx.doi.org/10.1088/0022-3727/32/12/311}
  {\path{doi:10.1088/0022-3727/32/12/311}}.

\bibitem{Segur2006/PSST}
P.~{S{\'e}gur}, A.~{Bourdon}, E.~{Marode}, D.~{Bessieres}, J.~H. {Paillol},
  {The use of an improved Eddington approximation to facilitate the calculation
  of photoionization in streamer discharges}, Plasma Sour. Sci. Technol. 15
  (2006) 648.
\newblock \href {http://dx.doi.org/10.1088/0963-0252/15/4/009}
  {\path{doi:10.1088/0963-0252/15/4/009}}.

\bibitem{Luque2007/ApPhL}
A.~{Luque}, U.~{Ebert}, C.~{Montijn}, W.~{Hundsdorfer}, {Photoionization in
  negative streamers: Fast computations and two propagation modes}, Appl. Phys.
  Lett. 90~(8) (2007) 081501.
\newblock \href {http://arxiv.org/abs/physics/0609247}
  {\path{arXiv:physics/0609247}}, \href {http://dx.doi.org/10.1063/1.2435934}
  {\path{doi:10.1063/1.2435934}}.

\bibitem{Liu2007/ApPhL}
N.~{Liu}, S.~{C{\'e}lestin}, A.~{Bourdon}, V.~P. {Pasko}, P.~{S{\'e}gur},
  E.~{Marode}, {Application of photoionization models based on radiative
  transfer and the Helmholtz equations to studies of streamers in weak electric
  fields}, Appl. Phys. Lett. 91~(21) (2007) 211501.
\newblock \href {http://dx.doi.org/10.1063/1.2816906}
  {\path{doi:10.1063/1.2816906}}.

\bibitem{Capeillere2008/JPhD}
J.~{Capeill{\`e}re}, P.~{S{\'e}gur}, A.~{Bourdon}, S.~{C{\'e}lestin},
  S.~{Pancheshnyi}, {The finite volume method solution of the radiative
  transfer equation for photon transport in non-thermal gas discharges:
  application to the calculation of photoionization in streamer discharges}, J.
  Phys. D 41~(23) (2008) 234018.
\newblock \href {http://dx.doi.org/10.1088/0022-3727/41/23/234018}
  {\path{doi:10.1088/0022-3727/41/23/234018}}.

\bibitem{Naidis1996/JPhD}
G.~V. {Naidis}, {On streamer interaction in a pulsed positive corona
  discharge}, J. Phys. D 29 (1996) 779.
\newblock \href {http://dx.doi.org/10.1088/0022-3727/29/3/039}
  {\path{doi:10.1088/0022-3727/29/3/039}}.

\bibitem{Luque2008/PhRvE}
A.~{Luque}, F.~{Brau}, U.~{Ebert}, {Saffman-Taylor streamers: Mutual finger
  interaction in spark formation}, \pre 78~(1) (2008) 016206.
\newblock \href {http://arxiv.org/abs/0708.1722} {\path{arXiv:0708.1722}},
  \href {http://dx.doi.org/10.1103/PhysRevE.78.016206}
  {\path{doi:10.1103/PhysRevE.78.016206}}.

\bibitem{Ratushna/prep}
V.~{Ratushnaya}, A.~{Luque}, U.~{Ebert}, {Electrodynamic characterization of
  long positive streamers in air} ({in preparation}).

\bibitem{Saffman1958/RSPSA}
P.~G. {Saffman}, G.~{Taylor}, {The Penetration of a Fluid into a Porous Medium
  or Hele-Shaw Cell Containing a More Viscous Liquid}, Royal Society of London
  Proceedings Series A 245 (1958) 312.
\newblock \href {http://dx.doi.org/10.1098/rspa.1958.0085}
  {\path{doi:10.1098/rspa.1958.0085}}.

\bibitem{Bensimon1986/RvMP}
D.~{Bensimon}, L.~P. {Kadanoff}, S.~{Liang}, B.~I. {Shraiman}, C.~{Tang},
  {Viscous flows in two dimensions}, Reviews of Modern Physics 58 (1986) 977.
\newblock \href {http://dx.doi.org/10.1103/RevModPhys.58.977}
  {\path{doi:10.1103/RevModPhys.58.977}}.

\bibitem{Meulenbroek2005/PhRvL}
B.~{Meulenbroek}, U.~{Ebert}, L.~{Sch{\"a}fer}, {Regularization of Moving
  Boundaries in a Laplacian Field by a Mixed Dirichlet-Neumann Boundary
  Condition: Exact Results}, Phys. Rev. Lett. 95~(19) (2005) 195004.
\newblock \href {http://arxiv.org/abs/nlin/0507019}
  {\path{arXiv:nlin/0507019}}, \href
  {http://dx.doi.org/10.1103/PhysRevLett.95.195004}
  {\path{doi:10.1103/PhysRevLett.95.195004}}.

\bibitem{Arrayas2002/PhRvL}
M.~{Array{\'a}s}, U.~{Ebert}, W.~{Hundsdorfer}, {Spontaneous Branching of
  Anode-Directed Streamers between Planar Electrodes}, Phys. Rev. Lett. 88~(17)
  (2002) 174502.
\newblock \href {http://arxiv.org/abs/nlin/0111043}
  {\path{arXiv:nlin/0111043}}, \href
  {http://dx.doi.org/10.1103/PhysRevLett.88.174502}
  {\path{doi:10.1103/PhysRevLett.88.174502}}.

\bibitem{Montijn2006/PhRvE}
C.~{Montijn}, U.~{Ebert}, W.~{Hundsdorfer}, {Numerical convergence of the
  branching time of negative streamers}, \pre 73~(6) (2006) 065401.
\newblock \href {http://arxiv.org/abs/physics/0604012}
  {\path{arXiv:physics/0604012}}, \href
  {http://dx.doi.org/10.1103/PhysRevE.73.065401}
  {\path{doi:10.1103/PhysRevE.73.065401}}.

\bibitem{Babaeva2008/ITPS}
N.~Y. {Babaeva}, M.~J. {Kushner}, {Streamer Branching: The Role of
  Inhomogeneities and Bubbles}, IEEE Trans. Plasma Sci. 36 (2008) 892.
\newblock \href {http://dx.doi.org/10.1109/TPS.2008.922434}
  {\path{doi:10.1109/TPS.2008.922434}}.

\bibitem{Naidis2010/JPhD}
G.~V. {Naidis}, {FAST TRACK COMMUNICATION: Modelling of streamer propagation in
  atmospheric-pressure helium plasma jets}, J. Phys. D 43 (2010) 2001.
\newblock \href {http://dx.doi.org/10.1088/0022-3727/43/40/402001}
  {\path{doi:10.1088/0022-3727/43/40/402001}}.

\bibitem{Franz1990/Sci}
R.~C. {Franz}, R.~J. {Nemzek}, J.~R. {Winckler}, {Television Image of a Large
  Upward Electrical Discharge Above a Thunderstorm System}, Science 249 (1990)
  48.
\newblock \href {http://dx.doi.org/10.1126/science.249.4964.48}
  {\path{doi:10.1126/science.249.4964.48}}.

\bibitem{Raizer1998/JPhD}
Y.~P. {Raizer}, G.~M. {Milikh}, M.~N. {Shneider}, S.~V. {Novakovski}, {Long
  streamers in the upper atmosphere above thundercloud}, J. Phys. D 31 (1998)
  3255.
\newblock \href {http://dx.doi.org/10.1088/0022-3727/31/22/014}
  {\path{doi:10.1088/0022-3727/31/22/014}}.

\bibitem{Pasko2010/JGRA}
V.~P. {Pasko}, {Recent advances in theory of transient luminous events}, J.
  Geophys. Res. (Space Phys) 115 (2010) 0.
\newblock \href {http://dx.doi.org/10.1029/2009JA014860}
  {\path{doi:10.1029/2009JA014860}}.

\bibitem{Luque2009/NatGe}
A.~{Luque}, U.~{Ebert}, {Emergence of sprite streamers from
  screening-ionization waves in the lower ionosphere}, Nature Geoscience 2
  (2009) 757.
\newblock \href {http://dx.doi.org/10.1038/ngeo662}
  {\path{doi:10.1038/ngeo662}}.

\bibitem{Luque2010/GeoRL}
A.~{Luque}, U.~{Ebert}, {Sprites in varying air density: Charge conservation,
  glowing negative trails and changing velocity}, Geophys. Res. Lett. 37 (2010)
  6806.
\newblock \href {http://dx.doi.org/10.1029/2009GL041982}
  {\path{doi:10.1029/2009GL041982}}.

\bibitem{Pasko1997/JGR}
V.~P. {Pasko}, U.~S. {Inan}, T.~F. {Bell}, Y.~N. {Taranenko}, {Sprites produced
  by quasi-electrostatic heating and ionization in the lower ionosphere}, \jgr
  102 (1997) 4529.
\newblock \href {http://dx.doi.org/10.1029/96JA03528}
  {\path{doi:10.1029/96JA03528}}.

\bibitem{Valdivia1997/GeoRL}
J.~A. {Valdivia}, G.~{Milikh}, K.~{Papadopoulos}, {Red sprites: Lightning as a
  fractal antenna}, Geophys. Res. Lett. 24 (1997) 3169.
\newblock \href {http://dx.doi.org/10.1029/97GL03188}
  {\path{doi:10.1029/97GL03188}}.

\bibitem{Marshall2010/JGRA/1}
R.~A. {Marshall}, R.~T. {Newsome}, N.~G. {Lehtinen}, N.~{Lavassar}, U.~S.
  {Inan}, {Optical signatures of radiation belt electron precipitation induced
  by ground-based VLF transmitters}, J. Geophys. Res. (Space Phys) 115 (2010)
  8206.
\newblock \href {http://dx.doi.org/10.1029/2010JA015394}
  {\path{doi:10.1029/2010JA015394}}.

\bibitem{Cummer2006/GeoRL}
S.~A. {Cummer}, N.~{Jaugey}, J.~{Li}, W.~A. {Lyons}, T.~E. {Nelson}, E.~A.
  {Gerken}, {Submillisecond imaging of sprite development and structure},
  Geophys. Res. Lett. 33 (2006) 4104.
\newblock \href {http://dx.doi.org/10.1029/2005GL024969}
  {\path{doi:10.1029/2005GL024969}}.

\bibitem{Stenbaek-Nielsen2008/JPhD}
H.~C. {Stenbaek-Nielsen}, M.~G. {McHarg}, {High time-resolution sprite imaging:
  observations and implications}, J. Phys. D 41~(23) (2008) 234009.
\newblock \href {http://dx.doi.org/10.1088/0022-3727/41/23/234009}
  {\path{doi:10.1088/0022-3727/41/23/234009}}.

\bibitem{Kanmae2007/GeoRL}
T.~{Kanmae}, H.~C. {Stenbaek-Nielsen}, M.~G. {McHarg}, {Altitude resolved
  sprite spectra with 3 ms temporal resolution}, Geophys. Res. Lett. 34 (2007)
  7810.
\newblock \href {http://dx.doi.org/10.1029/2006GL028608}
  {\path{doi:10.1029/2006GL028608}}.

\bibitem{Sentman2008/JGRD}
D.~D. {Sentman}, H.~C. {Stenbaek-Nielsen}, M.~G. {McHarg}, J.~S. {Morrill},
  {Correction to ``Plasma chemistry of sprite streamers''}, J. Geophys. Res.
  (Atmos.) 113 (2008) 14399.
\newblock \href {http://dx.doi.org/10.1029/2008JD010634}
  {\path{doi:10.1029/2008JD010634}}.

\bibitem{Liu2010/GeoRL}
N.~{Liu}, {Model of sprite luminous trail caused by increasing streamer
  current}, Geophys. Res. Lett. 37 (2010) 4102.
\newblock \href {http://dx.doi.org/10.1029/2009GL042214}
  {\path{doi:10.1029/2009GL042214}}.

\bibitem{Niemeyer1984/PhRvL}
L.~{Niemeyer}, L.~{Pietronero}, H.~J. {Wiesmann}, {Fractal Dimension of
  Dielectric Breakdown}, Phys. Rev. Lett. 52 (1984) 1033.
\newblock \href {http://dx.doi.org/10.1103/PhysRevLett.52.1033}
  {\path{doi:10.1103/PhysRevLett.52.1033}}.

\bibitem{Pasko2000/GeoRL}
V.~P. {Pasko}, U.~S. {Inan}, T.~F. {Bell}, {Fractal structure of sprites},
  Geophys. Res. Lett. 27 (2000) 497.
\newblock \href {http://dx.doi.org/10.1029/1999GL010749}
  {\path{doi:10.1029/1999GL010749}}.

\bibitem{Akyuz2003/JElec}
M.~Akyuz, A.~Larsson, V.~Cooray, G.~Strandberg, 3d simulations of streamer
  branching in air, J. Electrost. 59~(2) (2003) 115 -- 141.
\newblock \href {http://dx.doi.org/10.1016/S0304-3886(03)00066-4}
  {\path{doi:10.1016/S0304-3886(03)00066-4}}.

\end{thebibliography}
 
\end{document}